\documentclass[iop]{emulateapj}

\usepackage{natbib}
\usepackage[dvips]{color}
\usepackage{subfigure}

\def\e{\mathrm{e}} 
\newcommand{\rd}{\partial}

\newcommand{\bracket}[1]{\left<#1\right>}

\begin{document}

\title{On the importance of the equation of state for the
  neutrino-driven supernova explosion mechanism}

\author{
Yudai Suwa\altaffilmark{1},
Tomoya Takiwaki\altaffilmark{2},
Kei Kotake\altaffilmark{2,3}, 
Tobias Fischer\altaffilmark{4,5},
Matthias Liebend\"orfer\altaffilmark{6},
and Katsuhiko Sato\altaffilmark{7}
}

\altaffiltext{1}{Yukawa Institute for Theoretical Physics, Kyoto
  University, Oiwake-cho, Kitashirakawa, Sakyo-ku, Kyoto, 606-8502,
  Japan}
\altaffiltext{2}{Center for Computational Astrophysics, National
  Astronomical Observatory of Japan, Mitaka, Tokyo 181-8588, Japan}
\altaffiltext{3}{Division of Theoretical Astronomy, National
  Astronomical Observatory of Japan, Mitaka, Tokyo 181-8588, Japan}
\altaffiltext{4}{GSI, Helmholtzzentrum f\"ur Schwerionenforschung
  GmbH, Planckstr. 1, 64291 Darmstadt, Germany}
\altaffiltext{5}{Institut f{\"u}r Kernphysik, Technische
  Universit{\"a}t Darmstadt, Schlossgartenstra{\ss}e 9, 64289
  Darmstadt, Germany}
\altaffiltext{6}{Department of Physics, University of Basel,
  Klingelbergstr. 82, CH-4056 Basel, Switzerland} 
\altaffiltext{7}{The Institute for the Physics and Mathematics of the
  Universe, the University of Tokyo, Kashiwa, Chiba, 277-8568, Japan}

\begin{abstract}
By implementing widely-used equations of state (EOS) from Lattimer \&
Swesty (LS) and H. Shen et al. (SHEN) in core-collapse supernova
simulations, we explore possible impacts of these EOS on the
post-bounce dynamics prior to the onset of neutrino-driven
explosions. Our spherically symmetric (1D) and axially symmetric (2D)
models are based on neutrino radiation hydrodynamics including
spectral transport, which is solved by the isotropic diffusion source
approximation. We confirm that in 1D simulations neutrino-driven
explosions cannot be obtained for any of the employed EOS. Impacts of
the EOS on the post-bounce hydrodynamics are more clearly visible in
2D simulations.  In 2D models of a 15 $M_{\odot}$ progenitor using the
LS EOS, the stalled bounce shock expands to increasingly larger radii,
which is not the case using the SHEN EOS.  Keeping in mind that the
omission of the energy drain by heavy-lepton neutrinos in the present
scheme could facilitate explosions, we find that 2D models of an 11.2
$M_{\odot}$ progenitor produce neutrino-driven explosions for all the
EOS under investigation. Models using the LS EOS are slightly more
energetic compared to those with the SHEN EOS.  The more efficient
neutrino heating in the LS models coincides with a higher electron
antineutrino luminosity and a larger mass that is enclosed within the
gain region. The models based on the LS EOS also show a more vigorous
and aspherical downflow of accreting matter to the surface of the
protoneutron star (PNS). The accretion pattern is essential for the
production and strength of outgoing pressure waves, that can push in
turn the shock to larger radii and provide more favorable conditions
for the explosion. Based on our models we investigate several
diagnostic indicators of the explosion that have been suggested in the
literature, e.g., the amplitude of the standing-accretion shock
instability mode, the mass weighted average entropy in the gain
region, the PNS radius, the antesonic condition, the ratio of
advection and heating timescales, the neutrino heating efficiency, and
the growth parameter of convection.
\end{abstract}


\section{Introduction}
\label{sec:intro}

Core-collapse supernova explosions are triggered by the gravitational
energy released during the transition from a stellar core to a
protoneutron star (PNS). The thermodynamic conditions obtained at the
center of the stellar core are temperatures on the order of tens of
MeV and densities on the order of normal nuclear matter density
($3\times10^{14}$~g~cm$^{-3}$). There are only few equations of state
(EOS) available for these conditions. Several studies by
\citet{taka82} and \citet{baro85} were pioneering, investigating the
impact of the EOS on the explosion dynamics, albeit in a
phenomenological approach (see collective references in
\citealt{beth90}).  The most commonly used nuclear EOS in recent
supernova simulations are the EOS from \citet{latt91} (hereafter LS),
based on the incompressible liquid-drop model including surface
effects, and from \citet{shen98} (SHEN). The latter is based on
relativistic mean field (RMF) theory and the Thomas-Fermi
approximation. Recently, a new nuclear RMF EOS has been published in
\citet{hemp10}, which in addition can be obtained for several nuclear
parametrizations and provides detailed information about the chemical
composition (see also \citealt{gshen,furusawa}).

Thus, extensive studies have been performed to investigate the impact
of the nuclear EOS on supernova simulations. But most of these
simulations are limited to spherical symmetry. \citet{thom03}
investigated the effects of the incompressibility parameter on the
evolution of temperature and neutrino luminosities during the
postbounce evolution based on the three different incompressibilities
of the LS EOS ($K=180$, 220, 375~MeV). Only a small impact on the
dynamics was found. \citet{sumi06,sumi07} performed long-term
simulations of a 40 $M_\odot$ progenitor star from \citet{woos95}
comparing LS ($K=180$~MeV) and SHEN, focusing on the evolution until
black-hole formation and the emitted neutrino signal. \citet{fisc09}
included in addition progenitor stars from different stellar evolution
groups in the mass range of 40--50 $M_\odot$ and studied the neutrino
signal and collapse to a black hole in a systematic way. Their results
were confirmed and extended by \citet{ocon11} applying a simplified
treatment of neutrino transport. In \citet{hemp12}, the authors
applied their new EOS and systematically explored the hydrodynamic
evolution and emission of neutrinos for different nuclear parameters
within the class of RMF EOS. Moreover, \citet{sage09} and
\citet{fisc11} explored the appearance of quark matter via a first
order phase transition in supernova simulations during the early
post-bounce phase, which triggers the explosion even in spherically
symmetric models.

Furthermore, EOS studies based on multi-dimensional supernova models
have become available recently. \citet{kota04b} and \citet{sche10}
performed two-dimensional (2D) and three-dimensional (3D) simulations,
respectively, and investigated the initial post-bounce phase and
gravitational wave emission focusing on effects of rotation and
convection with simplified neutrino treatments. They found that the
gravitational wave signal depends strongly on the EOS (e.g.,
\citealt{kota11} for a recent review). In addition, \citet{mare09b}
investigated the EOS dependence of high frequency variations of the
neutrino luminosity and gravitational wave emission by performing 2D
simulations that include detailed neutrino-radiative transfer.

Several 2D studies obtain neutrino-driven explosions of massive
iron-core progenitors \citep[see
  e.g.][]{mare09b,brue09,suwa10,muel12b}, aided by the standing
accretion-shock instability (SASI) and neutrino-driven convection.
These simulations were evolved until the expanding explosion shock
left the central core ($\sim$1000~km) between about 400--600~ms after
bounce (depending on the progenitor and details of input
physics). Most multi-dimensional supernova studies used the LS EOS
with the incompressibility of $K=180$~MeV, which is a very soft EOS
\citep{bura06,suwa10}. In addition, \cite{mare09} performed 2D
simulation using the stiffer EOS from \citet{hill84}, and found that
the softer EOS results in a more optimistic situation for a possible
onset of an explosion. However, their simulation with the stiffer EOS
had been performed for a shorter post-bounce time than the softer
EOS. Hence, a final conclusion cannot be drawn.\footnote{Employing a
  stiffer EOS with $K=263$ MeV based on the Hartree-Fock
  approximation, Marek et al. found no explosions for the same
  progenitor model, whereas they indeed obtained an explosion for Shen
  EOS that is even stiffer with $K=281$ MeV \citep[see,
    e.g.,][]{jank12}.} Even further, they employed only two EOS, each
of which has a different incompressibility and symmetry energy.  Note
that the entire set of nuclear parameters, not only the
incompressibility and the symmetry energy, but also their density
dependence determine the resulting EOS.

In this paper we present results of numerical simulations of
core-collapse supernovae of massive iron-core progenitors. The spectral
neutrino transport is treated by the {\it Isotropic Diffusion Source 
 Approximation} (IDSA) \citep{lieb09}. We employ four EOS of LS (with the three
incompressibilities) and SHEN. We apply these different EOS in axially
symmetric (2D) simulations and investigate the differences obtained.
In the analysis of the data we focus on the post-bounce phase after the
shock-stall, on the neutrino-driven shock revival and the shock propagation 
 for more than 500 ms after core bounce.

The paper opens with the description of the the numerical method and
the employed EOS in Section 2. The results of 1D and 2D simulations
are presented in Section 3 and 4. We summarize our results and discuss
their implications in Section 5.


\section{Supernova Model} \label{sec:numerical}
\subsection{Hydrodynamics}

The basic evolution equations are written as follows,
\begin{equation}
 \frac{d\rho}{dt}+\rho\nabla\cdot\mathbf{v}=0,
\end{equation}
\begin{equation}
 \rho \frac{d\mathbf{v}}{dt}=-\nabla P -\rho \nabla \Phi
\end{equation}
\begin{equation}
\frac{\partial e^*}{\partial t}+
\nabla \cdot
\left[\left(e^* + P\right) \mathbf{v}\right]= -\rho \mathbf{v} \cdot \nabla \Phi+ Q_{\nu}\label{cool},
\end{equation}
\begin{equation}
 \bigtriangleup{\Phi} = 4\pi G \rho,
\end{equation}
with rest-mass density $\rho$, total pressure $P$ (including neutrino
contributions), matter velocity $\mathbf{v}$, total energy density
$e^*$, gravitational potential $\Phi$ as well as neutrino
heating/cooling rates $Q_{\nu}$. $\frac{d}{dt}$ is the Lagrangian time
derivative and $\frac{\partial}{\partial t}$ is the Eulerian time
derivative.

Our 2D simulations are performed using a code which is based on the
spectral neutrino transport scheme IDSA, developed by \cite{lieb09},
and the ZEUS-2D code \citep{ston92,suwa10,suwa11b}.\footnote{Our code
  was developed for core-collapse simulation and was used for the core
  collapse of very massive Population~III stars
  \citep{suwa07a,suwa07b,suwa09a}, where a leakage scheme for
  neutrinos was implemented \citep{kota03,taki04}.}  Details about the
IDSA are given in the next subsection.

In this work we use the conservation form for the total energy rather
than the internal energy equation used in the original ZEUS-2D code to
improve the accuracy of the total energy conservation. The source term
of the self-gravity in Eq. (\ref{cool}) is computed by the method
described in \cite{muel10}. The violation of the energy conservation
remains within 0.03\% of its gravitational binding energy ($\sim
10^{53}$ erg; see Appendix of \citealt{suwa11b}).
The simulations are performed on a grid of 300 logarithmically spaced
radial zones up to 5000 km and 128 equidistant angular zones covering
$0<\theta<\pi$. For neutrino transport, we use 20 logarithmically
spaced energy bins reaching from 3 to 300 MeV.

Our code is written in spherical coordinates, which have the following
advantage: We can compare 1D and 2D results of the ``same'' code. As
for 1D simulations, we just use ``one'' grid for the lateral
direction. This is impossible for a Cartesian grid code or a distorted
grid code such as VULCAN/2D \citep{burr06}.
In fact, results obtained in 1D and 2D simulations show exactly the
same features for the preshocked region \cite[see Figure 1
  of][]{suwa10}. Furthermore, using the detailed comparison study of,
e.g., \cite{lieb05} based on spherically symmetric simulations, it is
possible to validate our code \citep[see Appendix in][]{suwa11b}.
Therefore, we can directly identify the multi-dimensional effects by
using this code both in 1D and 2D simulations. On the other hand, the
spherical coordinates have a disadvantage for the CFL condition
because the length of the lateral direction at the center becomes much
smaller than the length of the radial direction, which can be written
as $r_1\Delta\theta\sim0.02 r_1(128/N_\theta)$. Here, $r_1$,
$\Delta\theta$, and $N_\theta$ are the mesh width of the radial
direction at the center, that of the latitude, and the mesh number of
the lateral direction. This implies that the time step of a 2D
simulation with this lateral resolution must be shortened by a factor
of 50 compared to a 1D simulation. Therefore, in order to perform
long-term simulations in spherical coordinates, we should employ some
sort of approximation. In this study, the innermost core of $\sim$5 km
radius is computed in spherical symmetry, similar to what was done in
\cite{bura06}.

\subsection{Neutrino transfer}
In this subsection, we describe the scheme of the neutrino transfer
used in this study.  This scheme is based on the description in
\citet{lieb09}.

We solve the transport equation for neutrino distribution function
$f_l(t,r,\mu,E)$ depending on the time $t$, radius $r$, and the
momentum phase space spanned by the angle cosine $\mu$, of the
neutrino propagation direction with respect to the radius, and the
neutrino energy $E$. In this calculation, we include two neutrino
species ($l=\nu_\e, \bar\nu_\e$).  We divide the distribution function
of neutrinos into two parts as
\begin{equation}
f_l=f_l^t+f_l^s,
\end{equation}
where $f^t$ is an isotropic distribution function of trapped particles
and $f^s$ is a distribution function of streaming particles,
representing neutrinos of a given species and energy which find the
local zone opaque or transparent, respectively. We solve each
component separately as
\begin{eqnarray}
&&D(f_l^t)=j_l-(j_l+\chi_l)f_l^t-\Sigma_l,\label{eq:ft}\\
&&D(f_l^s)=   -(j_l+\chi_l)f_l^s+\Sigma_l,
\end{eqnarray}
where $D()$ is an operator describing particle propagation, $j$ is a
particle emissivity, $\chi$ is a particle absorptivity, and $\Sigma$
is the {\it diffusion source}, which converts trapped particles into
streaming particles and vice versa. In this approximation, $\Sigma$ is
determined by the diffusion limit (see \citealt{lieb09} for detail).
Weak interactions are implemented as described in \cite{brue85}.  We
assume that the spectrum of $f^t$ is thermal and treat trapped
particles as a fluid element. We characterize the thermal equilibrium
by a particle number fraction, $Y^t$, and a particle mean specific
energy, $Z^t$,
\begin{eqnarray}
Y_l^t=\frac{m_\mathrm{b}}{\rho}\frac{4\pi}{(hc)^3}\int f_l^t \epsilon_\nu^2d\epsilon_\nu,
\label{eq:yt}\\
Z_l^t=\frac{m_\mathrm{b}}{\rho}\frac{4\pi}{(hc)^3}\int f_l^t \epsilon_\nu^3d\epsilon_\nu,
\label{eq:zt}
\end{eqnarray}
where $m_\mathrm{b}$ is the baryon mass, $\rho$ is the baryon density,
$h$ is the Planck constant, $c$ is the speed of light, and
$\epsilon_\nu$ is the energy of neutrinos.  The advection of these
values is calculated in the same way as hydrodynamic quantities.
Using the net interaction rates, $s_l=j_l-(j_l+\chi_l)(f_l^t+f_l^s)$,
between matter and neutrinos, we can write the following changes of
the electron fraction, $Y_\e$ and the internal energy, $e$:
\begin{eqnarray}
\frac{\rd Y_\e}{\rd t}=-\frac{m_b}{\rho}\frac{4\pi c^2}{(hc)^3}\int (s_{\nu_\e}-s_{\bar\nu_e})\epsilon_\nu^2 d\epsilon_\nu,\label{eq:dyedt}\\
\frac{\rd e}{\rd t}=-\frac{m_b}{\rho}\frac{4\pi c^2}{(hc)^3}\int (s_{\nu_\e}+s_{\bar\nu_e})\epsilon_\nu^3 d\epsilon_\nu.\label{eq:dedt}
\end{eqnarray}
The changes of the electron fraction and specific internal energy feed
back the emissivity and absorptivity into Eq. (\ref{eq:ft}), so that
we should find the consistent solution by iterating Eqs.(\ref{eq:ft}),
(\ref{eq:dyedt}), and (\ref{eq:dedt}). Once the consistent solution
has been found, Eq. (\ref{eq:ft}) gives the updated distribution
function and specific energy of the trapped particles, which leads to
updated $Y_l$ and $Z_l$ values by Eqs. (\ref{eq:yt}) and
(\ref{eq:zt}).

\subsection{Equation of State}
\label{sec:eos}

The equation of state (EOS) in supernova simulations needs to handle a
large variety of thermodynamic conditions that reflect different
nuclear regimes. For temperatures below 0.4~MeV\footnote{Although the
  threshold temperature is slightly higher (around 0.5-0.8 MeV,
  depending on density and isospin asymmetry), we have tested also the
  higher flash temperatures and found no significant difference.},
time-dependent nuclear reactions are important and the baryon
composition is dominated by heavy nuclei. Above this temperature
nuclei are in chemical and thermal equilibrium, known as nuclear
statistical equilibrium (NSE), where the baryon composition is given
by temperature, baryon density and proton-to-baryon ratio (or
equivalently the electron fraction in the absence of, e.g., muons). At
high densities, typically above normal nuclear matter density, the
transition to a state of matter composed of deconfined quarks may be
possible. In addition to the baryons (or quarks), a gas of degenerate
and possibly relativistic electrons is present. Also (non-degenerate)
positrons and photons must be taken into account.

In this study we employ in supernova simulations the two standard EOS
for matter in NSE, the \citet{latt91} EOS (LS) and the \citet{shen98}
EOS (SHEN).
LS is based on the incompressible liquid drop model for nuclei,
combined with dripped nucleons. The transition to uniform nuclear
matter is calculated via a Maxwell phase transition. Contributions
from electrons/positrons and photons are taken into account
intrinsically. LS has a symmetry energy of 29.3~MeV. Three different
values of the incompressibility parameter are available, $K=180$~MeV
(LS180), 220~MeV (LS200), and 375~MeV (LS375). With these one obtains
maximum neutron star masses of $\sim1.8 M_\odot$ (LS180), $\sim 2.0
M_\odot$ (LS220), $\sim2.7 M_\odot$ (LS375) \citep[see][]{ocon11}.

SHEN gives only contributions for baryons valid for matter in NSE. It
is based on the relativistic mean field theory and the Thomas-Fermi
approximation for the description of heavy nuclei.  The symmetry
energy and the incompressibility are 36.9~MeV and $K=281$~MeV,
respectively. It results in a maximum neutron star mass of $\sim2.2
M_\odot$ \citep[see][]{ocon11,kiuchi08}.

Although LS180 fails to fulfill the recent neutron-star mass
constraint from \citet{demo10} of $1.97\pm0.05 M_\odot$, we
nevertheless include it in our EOS comparison study here because it is
widely used and because of its similarity to LS220, which does fulfill
the Demorest~\emph{et~al.} constraint.

There are only few experimental constraints on high-density and
finite-temperature nuclear EOS. For zero-temperature symmetric nuclear
matter, composed of equal numbers of protons and neutrons, the
symmetry energy can be well determined.  Moreover, recent constraints
from nuclear theory, experiments and observations predict a symmetry
energy between 31--33.6~MeV with a density gradient between 49.1 and
80~MeV \citep[see][]{latt12}.  These parameters produce cold neutron
star radii between 11--12~km, in agreement with the results of recent
efforts to consistently include three-body forces based on low-energy
chiral perturbation theory \citep[see, e.g.,][]{stei10,hebe10}.
Heavy-ion collision experiments can only hardly be used to constrain
the supernova EOS, because the conditions in the collisons are rather
isospin-symmetric such that they only cover slightly neutron-rich
conditions where $Y_e\simeq0.485$. But matter in supernova interiors
can be extremely neutron rich with $Y_e=0.01$--0.3.  Furthermore, the
incompressibility constraint of 200~MeV~$\lesssim K \lesssim$~300~MeV
is still weak.
 
In this paper, we investigate the impact of the EOS on the dynamics of
core-collapse supernovae in 1D and 2D simulations. Since the employed
EOS cover a wide range of nuclear parameters, we can identify the
systematic features that are related to the different EOS.

\section{Simulations in spherical symmetry}\label{sec:1d}

In the following subsections we compare results obtained in
simulations using the four EOS which were discussed in
\S\ref{sec:eos}. We use a 15 $M_\odot$ progenitor from
\cite{woos95}. We will start with spherically symmetric simulations
focusing on the differences between the different LS EOS. Further
below, we will also discuss results obtained using SHEN.

\begin{figure}[tbp]
\includegraphics[width=0.45\textwidth]{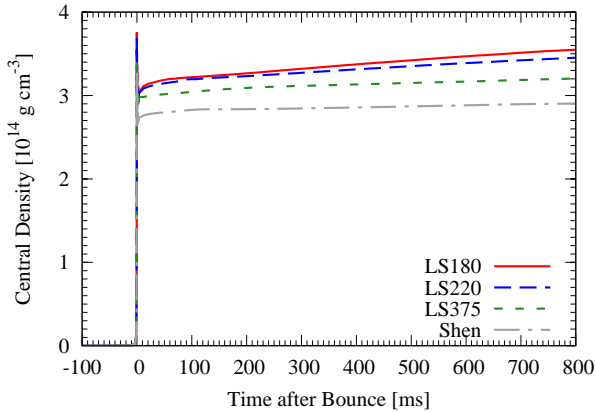}
\caption{Evolution of the central density, comparing the different EOS
  under investigation (LS180: red solid line, LS220: blue dashed line,
  LS375: green dotted line, and SHEN: gray dot-dashed line).}
\label{fig:cden}
\end{figure}

The supernova evolution can be separated into the following phases;
core collapse, bounce, prompt shock propagation, neutrino burst, and
accretion phase \citep{kota06,jank07}. Note that for the post-bounce
times considered, explosions could not be obtained in spherical
symmetry \citep[see][for more details]{suwa10,suwa11b}.

\begin{figure*}[tbp]
\centering
\subfigure[Velocity]{\includegraphics[width=0.475\textwidth]{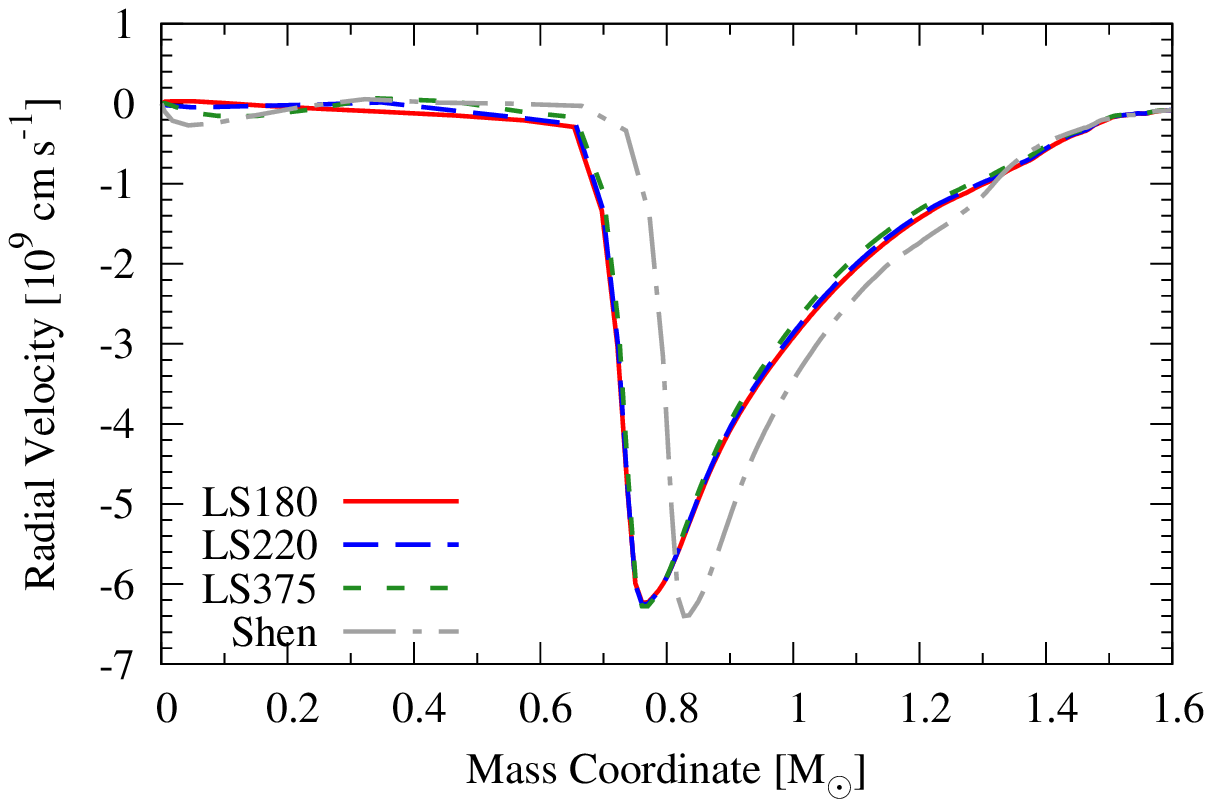}}
\subfigure[Electron fraction]{\includegraphics[width=0.475\textwidth]{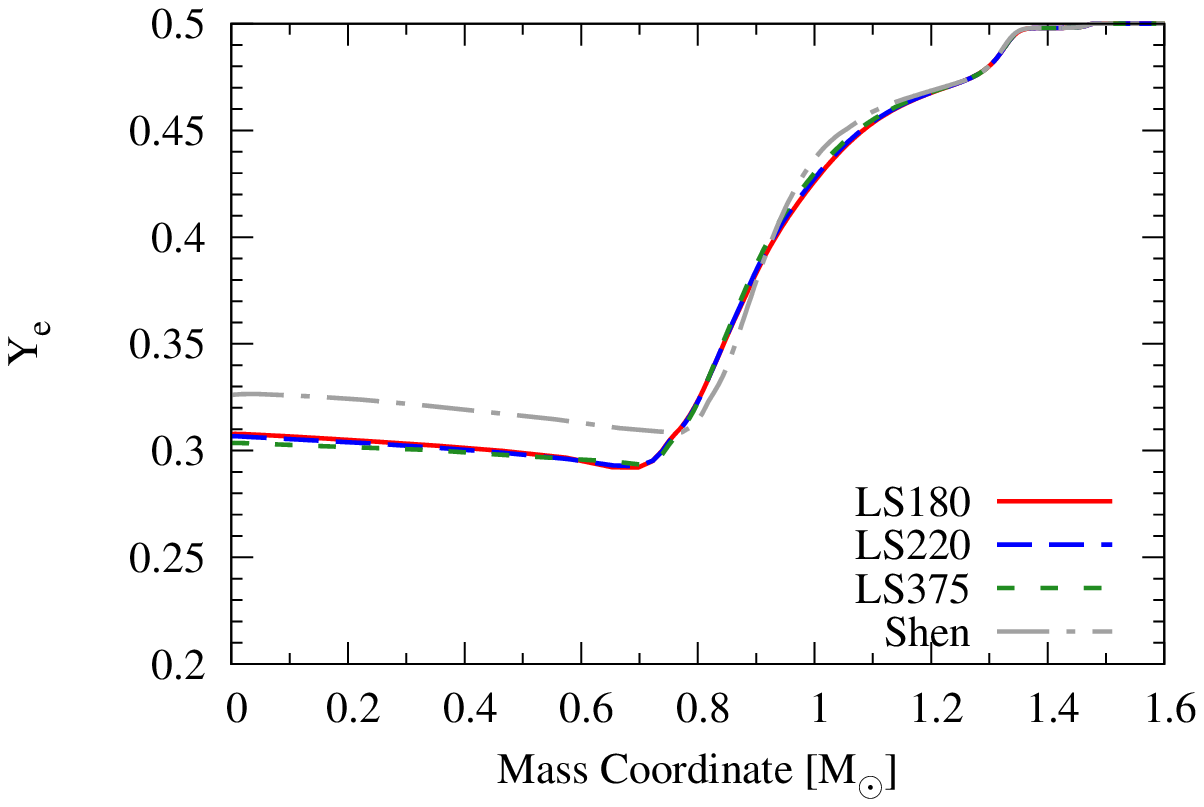}}
\subfigure[Temperature]{\includegraphics[width=0.475\textwidth]{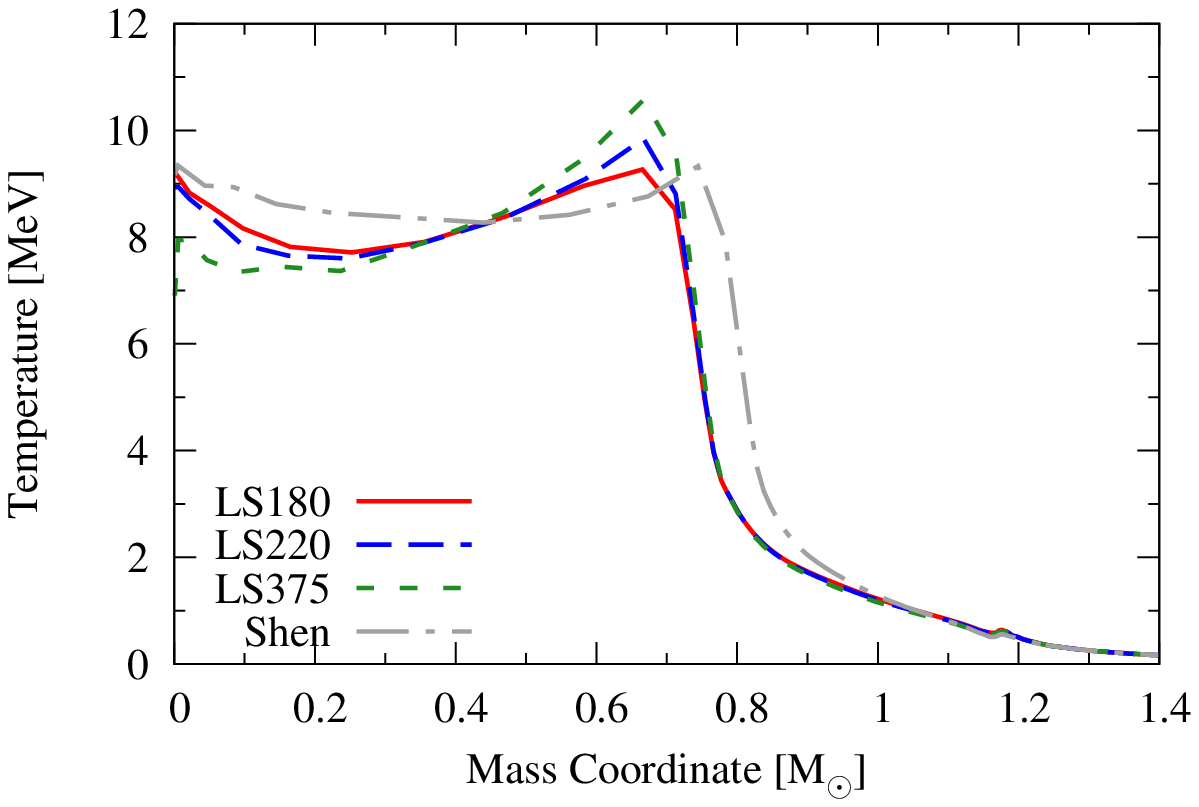}}
\subfigure[Entropy per baryon]{\includegraphics[width=0.475\textwidth]{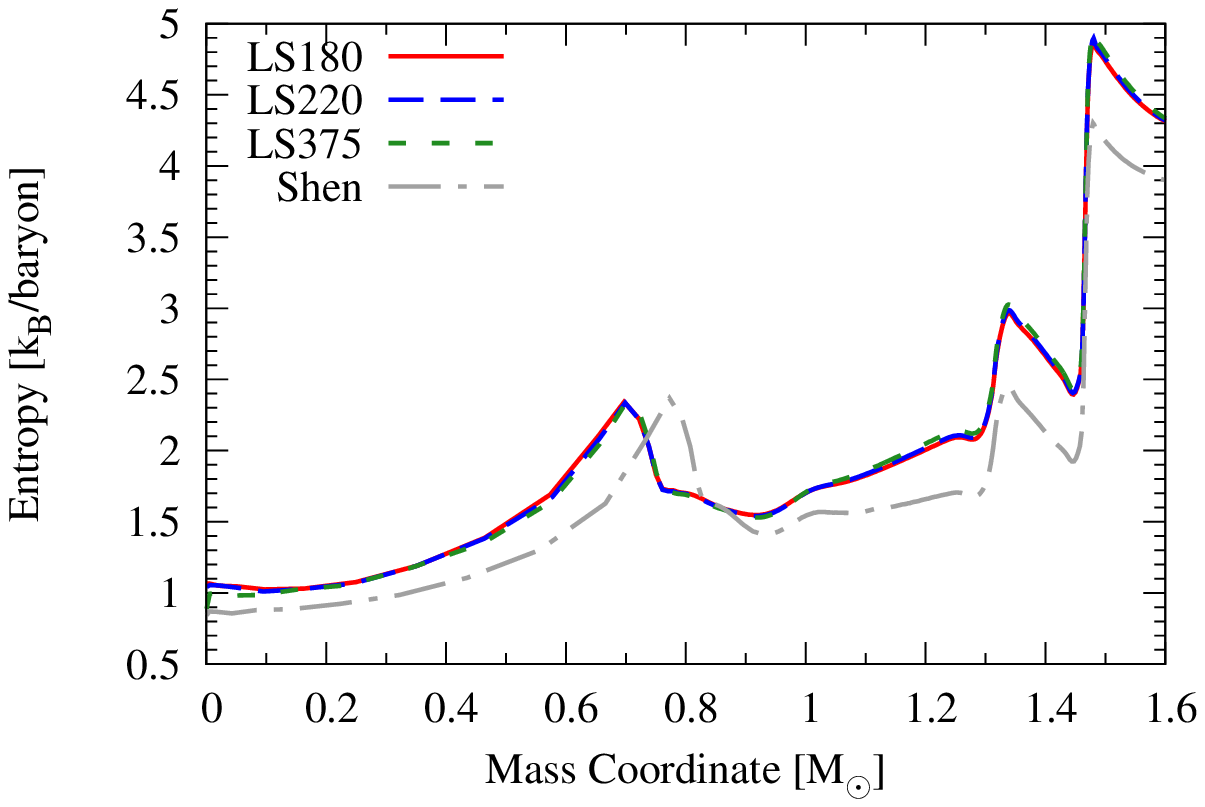}}
\caption{Radial bounce profiles as a function of the enclosed baryon
  mass. The velocity jump (at $\sim 0.7M_\odot$ for LS EOS and $\sim
  0.8M_\odot$ for SHEN) mark the edge of the inner core, above which
  the accretion flow is supersonic. The inner core at this time is not
  swept by the shock wave so that the inner core maintains its low
  entropy.}
\label{fig:bounce}
\end{figure*}

\subsection{Comparison between different LS EOS}

Figure \ref{fig:cden} shows the time evolutions of the central
density.  Because of the difference of the incompressibility $K$
(large $K$ implies a stiff EOS for otherwise identical nuclear
parameters), LS180 (red solid line) results in the highest central
density, while LS375 (green dotted line) corresponds to the lowest.
However, the central density differences between these EOS is
$\lesssim$ 10\%. Since LS220 exhibits almost identical features as
LS180 during the post-bounce phase up to about 500 ms and because
LS180 has been used in many previous publications of supernova
simulations, we mainly compare LS180 and LS375 in the following.
Figure \ref{fig:bounce}, shows the radial bounce profiles of velocity,
$Y_e$, temperature and entropy per baryon as a function of the
enclosed mass. The slightly different temperatures obtained right
behind the shock and at the very center for the different LS EOS, are
due to the different incompressibilities.  Highest (lowest)
temperatures behind the shock, as well as lowest (highest)
temperatures at the stellar core, are reached using LS375 (LS180) (see
Figure~\ref{fig:bounce}). Moreover, the lowest (highest) $Y_e$ is
obtained using LS375 (LS180). However, the differences obtained are
small and all LS EOS have very similar profiles.  Thus, despite the
difference of the central density, the structures of the inner cores
at bounce are very similar for all LS EOS, especially the mass
enclosed inside the bounce shock and hence the energetics of the
bounce shock is practically the same.

\begin{figure*}[tbp]
\centering
\subfigure[Shock evolution]{\includegraphics[width=\columnwidth]{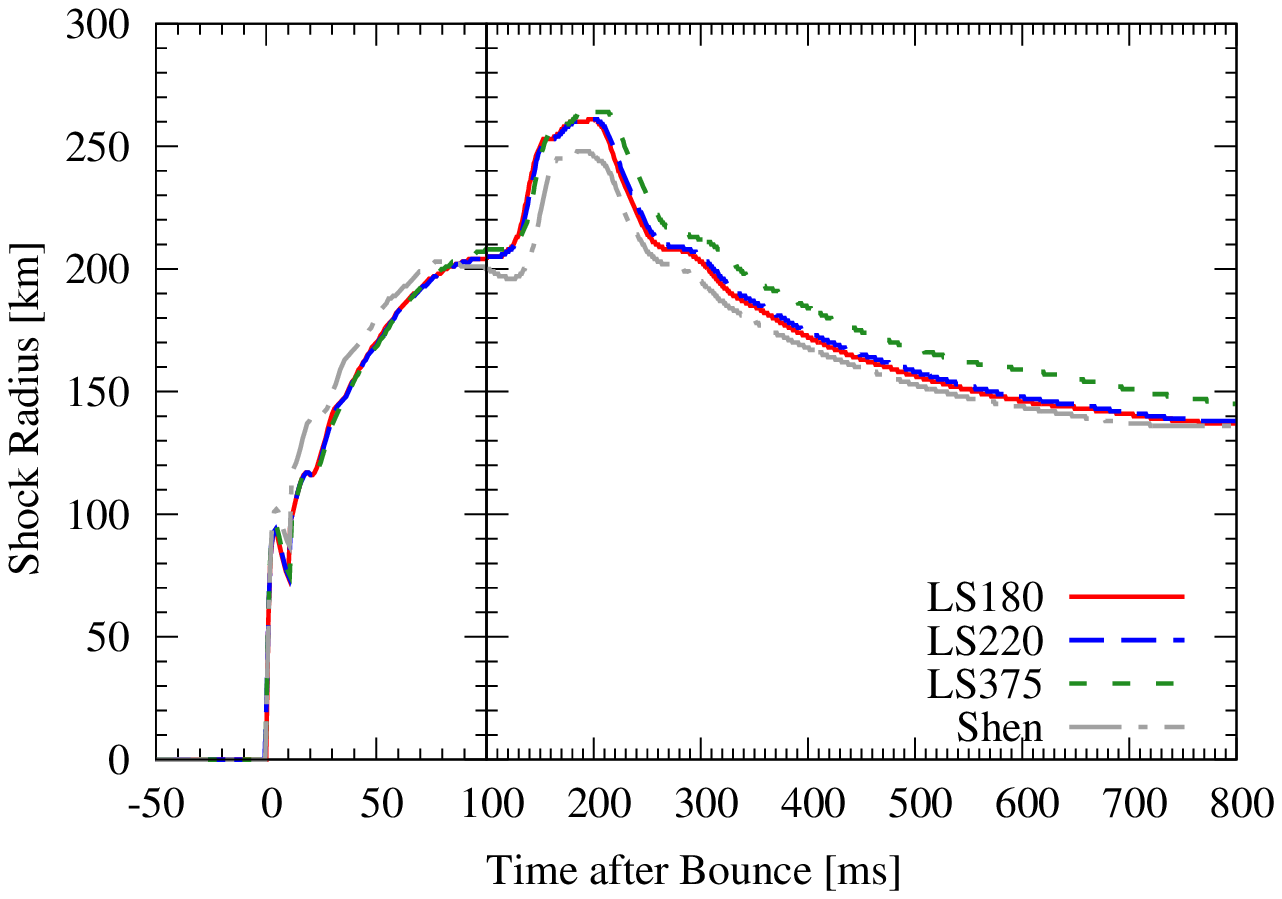}
\label{fig:shock}}
\hfill
\subfigure[Electron fraction evolution]
{\includegraphics[width=\columnwidth]{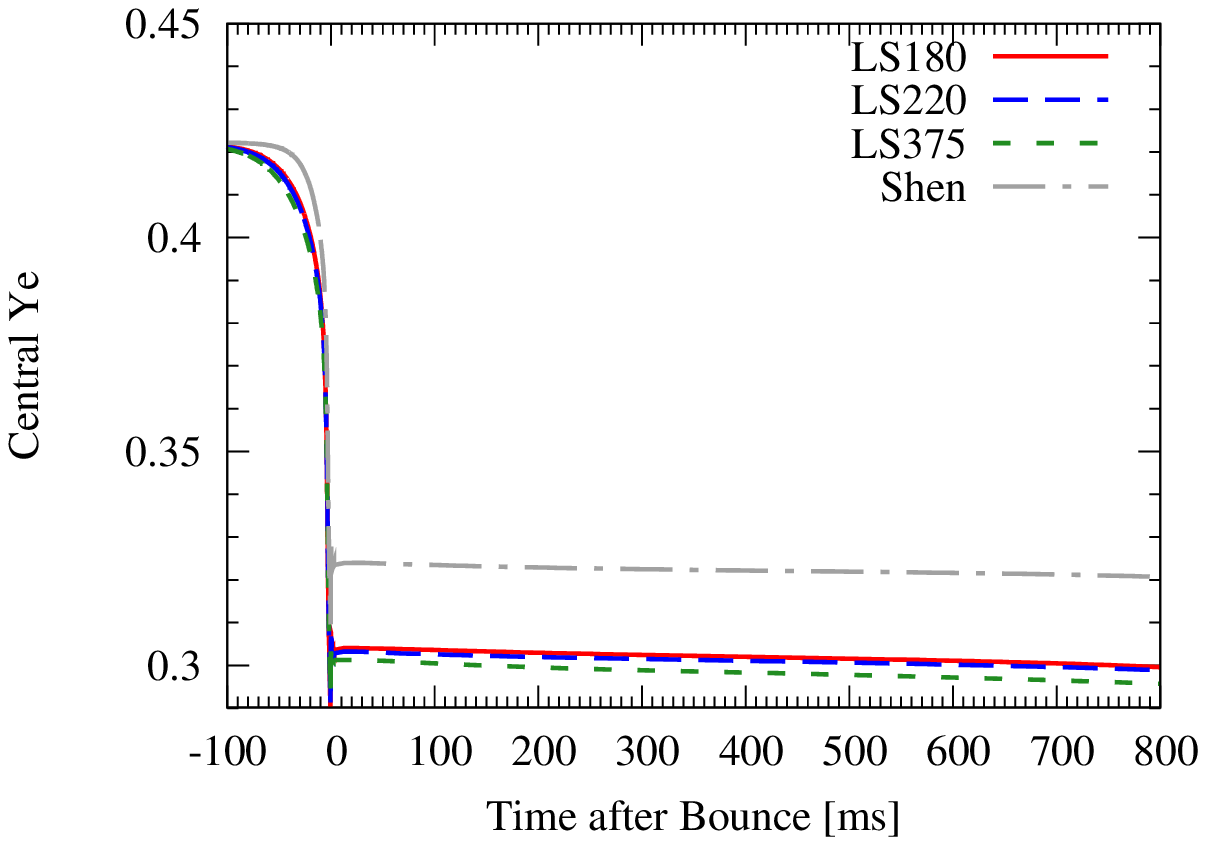}
\label{fig:cye}}
\caption{The left panel shows the evolution of the shock radius. Note
  that the time axis uses two different scales to clearly show the
  early post-bounce phase.  The right panel shows the evolution of the
  central electron fraction for the different EOS under
  investigation.}
\end{figure*}

All LS EOS show a similar shock evolution post bounce up to about
200~ms post bounce, expanding to $\sim$ 250~km. After that, the shocks
stall and turn into standing accretion shocks. After about $\sim
200$~ms post bounce, LS375 has a larger shock radius than LS180 and
LS220 (see Figure \ref{fig:shock}).

\subsection{Comparison between LS and SHEN EOS}

\begin{figure}[tbp]
\includegraphics[width=0.45\textwidth]{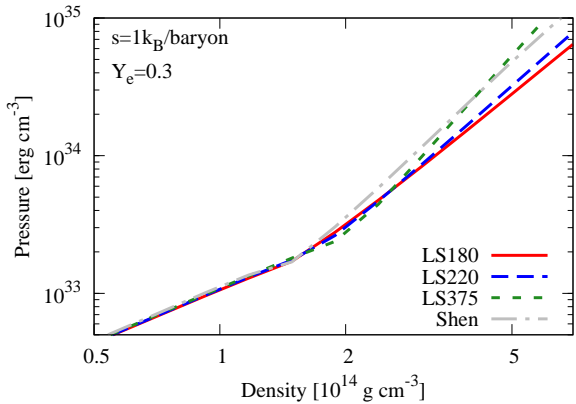}
\caption{The pressure as function of the density for the case of
  $s$=1$\mathrm{k_B}/$baryon and $Y_e=0.3$. These conditions
  correspond to the most central part of the collapsing core where the
  neutrinos are completely trapped. Above nuclear density ($\approx
  3\times 10^{14}$) g cm$^{-3}$ the difference between the EOS is
  significant.}
\label{fig:dp_s1}
\end{figure}

SHEN exhibits a significantly lower central density compared to LS375
which is at first view counter-intuitive because SHEN has a lower
incompressibility ($K=281$~MeV) than LS375 ($K=375$~MeV).  This aspect
can be related to the larger adiabatic index of SHEN at high densities
around $\approx 2\times10^{14}$ g cm$^{-3}$. It means that there are
regimes where curves of constant entropy per baryon in a $\rho-p$
diagram have a steeper gradient for SHEN.  Figure \ref{fig:dp_s1}
shows the pressure as a function of the density for the employed EOS,
at fixed entropy per baryon of 1~k$_B$ and electron fraction of
$Y_e=0.3$.  Only at very high densities, $\gtrsim 3\times 10^{14}$ g
cm$^{-3}$, LS375 has a steeper density gradient and becomes stiffer
than SHEN, mainly due to the higher $K$. This example illustrates that
neither incompressibility nor symmetry energy alone determine the
nuclear properties of the EOS.  It is rather the entire ensemble of
nuclear parameters, and in particular also the gradients of the
incompressibility and symmetry energy, which determine the EOS and
hence the possible outcome of dynamical simulations.

At core bounce and shortly after, the shock wave propagates at high
densities, where the highest pressure of SHEN results in the largest
shock radius. Later during the post-bounce evolution, the bounce shock
propagates towards lower densities on the order of $10^{9-13}$~g
cm$^{-3}$, where all EOS are very similar (Figure \ref{fig:shock}).

Bounce conditions of selected quantities are illustrated in
Figure~\ref{fig:bounce} in comparison to the LS EOS. The most
important difference is the larger mass enclosed inside the core at
bounce for SHEN. This difference cannot only be related to the
incompressibility. More relevant is the symmetry energy, which mainly
affects the evolution of the electron fraction $Y_e$. Note that the
dominant contributions to the pressure are given by the degenerate
electron gas during the core-collapse phase, while after bounce the
dominating contributions come from nucleons in the highly dissociated
regime behind the shock wave. Figure~\ref{fig:cye} shows the evolution
of the central $Y_e$ for the different EOS. $Y_e$ is very similar for
all LS EOS. But for SHEN, $Y_e$ is significantly larger than for any
LS EOS, already during core collapse. At bounce this difference
results in $Y_e\simeq0.32$ for SHEN and $Y_e\simeq0.3$ for LS.  It
indicates a slower iron-core deleptonization during collapse and
explains the larger core mass at bounce of about 0.8 $M_\odot$ for
SHEN. The models based on LS EOS result in core masses of about 0.7
$M_\odot$.  The different shock positions at bounce relate to these
differences in the structures of the core (see the temperature and
$Y_e$ profiles in Figure~\ref{fig:bounce}) and hence reflect the
different shock energetics during the initial propagation. Note the
lower central entropy per baryon for SHEN in comparison to LS, which
is mainly due to the well known problem of the interactive LS EOS
implementation, producing a systematically too low fraction of
$\alpha$-particles at low density. We did not correct for this in the
current study. Note also the lower entropy per baryon for SHEN outside
about 1.5 $M_\odot$. It is related to the different non-NSE
treatments, based on the ideal Si-gas approximation, in LS and
SHEN. Our findings are consistent with previous studies comparing LS
and SHEN, see e.g. \citet{sumi05} and \citet{hemp12}.  It is also
worth mentioning that the incorrect alpha-particle fraction of LS EOS
has a negligible impact on neutrino heating/cooling in the post-shock
regions because all nuclei (including alpha-particles) dissociate into
free nucleons (e.g., panel (e) of Figure 7 in
\citealt{hemp12}). Moreover in the preshock regions, inelastic
neutrino scattering of light nuclei, which was proposed to preheat
material there (\citealt{haxton88}, see \citealt{ohnishi08} for
collective references therein), has recently been shown to have
negligible impact on assisting the neutrino-driven explosions
\citep{langanke08}.

Note that the inner-core mass of 0.7 and 0.8 $M_\odot$ for the
enclosed baryon mass at bounce are too high values that are due to the
approximations employed in our simulations.  We omit neutrino
scattering on electrons/positrons and general-relativistic effects,
which lead to smaller inner core masses at bounce (see
\citealt{lieb01,thom03}).  In addition, we use the electron-capture
rates from \citet{brue85}, also for heavy nuclei, which assume a
single representative nucleus with average charge and mass. Improved
electron-capture rates on heavy nuclei from \citet{lang03}, which are
based on the NSE distribution of heavy nuclei, also lead to smaller
core masses at bounce (see \citealt{hix03} and \citealt{jank07} for
details).

\begin{figure}[tbp]
\includegraphics[width=0.45\textwidth]{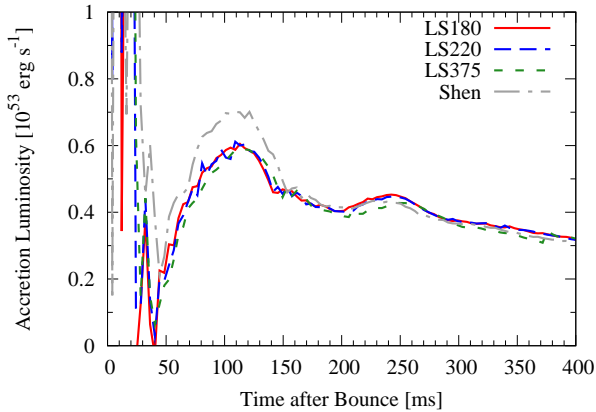}
\caption{Post-bounce evolution of the accretion luminosity for the EOS
  under investigation.}
\label{fig:acc_lum}
\end{figure}

These size differences of the inner core mass lead to different shock
evolutions. Due to the larger inner core, SHEN initially shows a
slightly more rapid expansion of the shock wave.  However, $\sim
50$~ms after bounce the shock contracts faster towards the PNS for
SHEN than for any LS EOS.  The reason is a larger energy loss with
SHEN when the shock propagates with its larger energy across the
neutrinospheres, from where the deleptonization burst is released.
This effect can also be seen in the luminosities. Before the explosion
is launched, the electron (anti)neutrino luminosity is determined by
mass accretion during the post-bounce evolution. With the accretion
rate $\dot{M}$ and an enclosed mass $M$ at the neutrinospheres $R_\nu$
one obtains the accretion luminosity as
\begin{equation}
L_\mathrm{acc} \simeq 8\times10^{52}
\left(\frac{M}{1.5~M_\odot}\right)
\left(\frac{50~\text{km}}{R_\nu}\right)
\left(\frac{\dot{M}}{1~M_\odot\text{s}^{-1}}\right)
\text{erg~s}^{-1}.
\end{equation}
Figure \ref{fig:acc_lum} shows the post-bounce evolution of this
quantity for the different EOS under investigation. The highest mass
accretion rates for SHEN explain the highest luminosities for SHEN
until about 150~ms post bounce, in comparison to the LS EOS. During
the later post-bounce evolution, the mass accretion rates at the
neutrinospheres become very similar for all EOS considered. Only small
differences can be identified with slightly lower luminosities for the
stiffer LS375 and SHEN on the one hand, and slightly higher
luminosities for the softer LS180 and LS220 on the other hand.

\section{Results of 2D simulations}\label{sec:2d}

In this section, we present our results of 2D simulations. We perform
2D simulations for LS180, LS375, and SHEN. Since LS220 shows almost
the same features as LS180 as seen in the previous section, we do not
include it in this section.

\subsection{Comparison between 1D and 2D using LS180}

\begin{figure}[tbb]
\centering
\includegraphics[width=0.45\textwidth]{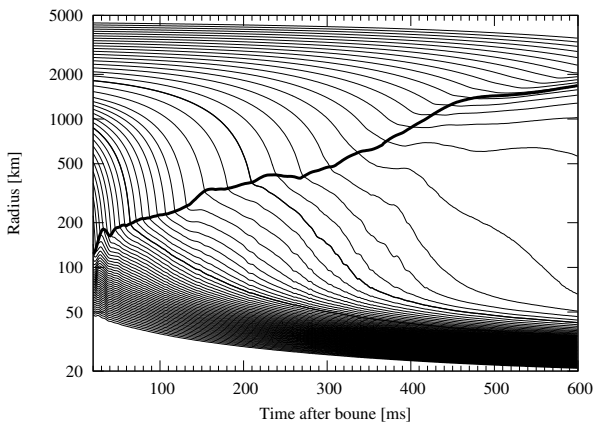}
\caption{The mass trajectory as a function of time after the bounce
  for 2D simulation of 15 $M_\odot$ with LS180.  The thick black line
  represents the angle-averaged shock radius.  The grey lines show the
  mass from 1.0 to 1.7 $M_\odot$ at intervals of 0.01 $M_\odot$.  Two
  thin black lines indicate 1.4 and 1.5 $M_\odot$, respectively.}
\label{fig:mass_ww15}
\end{figure}

First of all, we compare the results of spherically symmetric and
axially symmetric simulations using LS180. For the 2D simulation we
obtain a weak explosion, corresponding to the so-called ``passive
expansion''\citep{bura06,suwa10}, in which the rate of mass accretion
through the shock is larger than that of mass-ejection, so that the
mass of the central PNS is still increasing after the onset of shock
expansion\footnote{Note that the post-shock material have partly
  positive velocities (e.g., Figure 9(a)).}.  On average, only the
continuous shock expansion to larger radii is obtained with yet no
significant matter outflow (see Figure \ref{fig:mass_ww15}) for the
simulation times considered. The shock expansion is an effect of the
accumulation of hot matter in combination with a decreasing ram
pressure of the material ahead of the shock. The ram pressure
decreases because of a decreasing accretion rate and the propagation
of the shock to larger radii where the gravitational potential is less
deep.

\begin{figure*}[htbb]
\centering
\includegraphics[width=1.\textwidth]{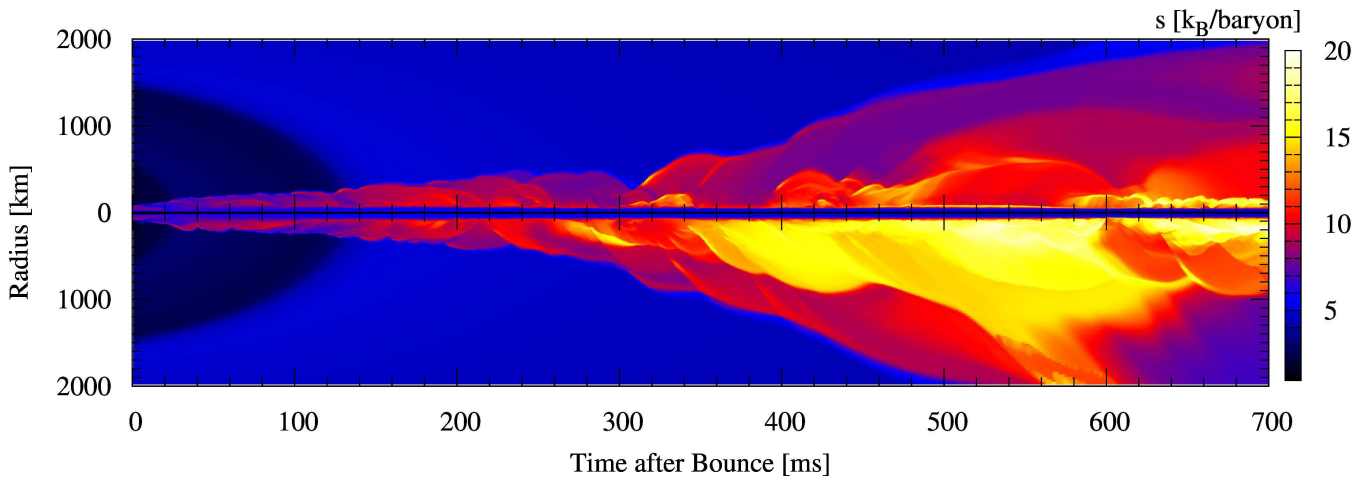}
\caption{Post-bounce evolution of the entropy distribution,
  illustrated in a space-time diagram. The top and bottom panels
  correspond to the northern pole and southern pole, respectively. The
  high-entropy region (red or yellow) shows the post shock layer. The
  shock wave oscillation produced by SASI and convective motion is
  apparent. The shock wave at the southern pole reaches a distance of
  2000 km from the center earlier than at the north pole,
  corresponding to the existence of a $\ell=1$ mode of the SASI.}
\label{fig:entropy_pole}
\end{figure*}

In Figure~\ref{fig:entropy_pole} we show the entropy distribution at
the north pole (top panel) and the south pole (bottom panel) for the
2D simulation. The deviation from a spherically symmetric distribution
becomes prominent after about 20~ms post bounce when convection starts
behind the shock and shock wave oscillations start to introduce
substantial differences of the shock position at the north and south
poles. The standing accretion shock instability (SASI) drives such
shock oscillations on a longer timescale on the order of several
100~ms. Until about 300~ms post bounce, the standing accretion shock
expands only slowly to several 100~km. After that, the shock expands
faster and reaches about 1000~km at the north pole and 2000~km at the
south pole at about 500~ms post bounce.  We will return to the origin
of the increasing difference between north- and south-pole evolution
further below.  Here, we will discuss the entropy distribution
obtained in comparison with the spherically symmetric simulation.

\begin{figure*}[tbb]
\centering
\includegraphics[width=1.\textwidth]{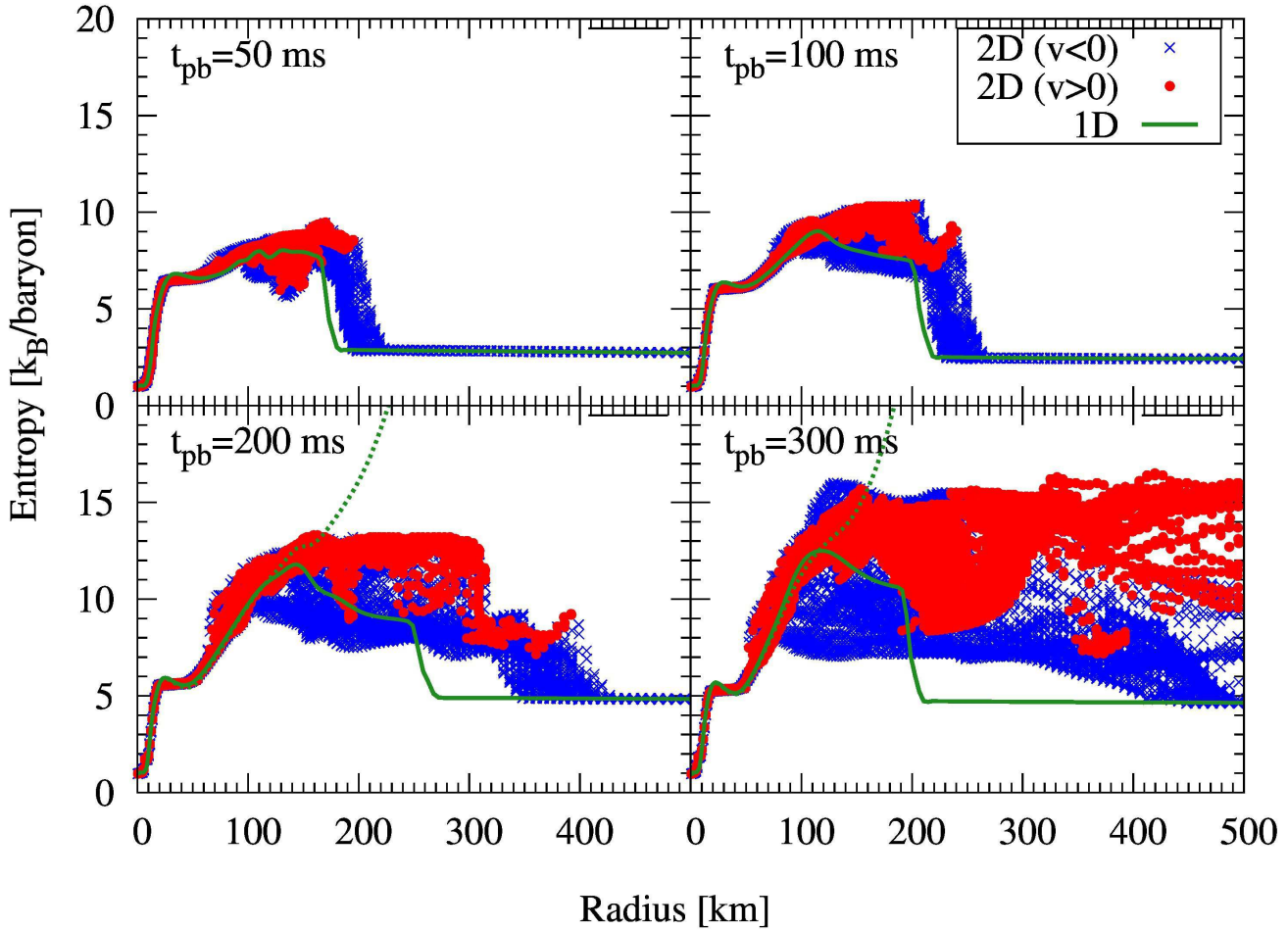}
\caption{Radial entropy distribution of the simulations with
  LS180. Red and blue points represent result of the 2D simulation,
  the green solid line corresponds to the 1D results. In addition, the
  green dotted line shows the entropy in weak equilibrium that would
  be achieved by infinitely-long exposure of the matter to the
  prevailing neutrino abundances and spectra.}
\label{fig:entropy_2d}
\end{figure*}

Figure~\ref{fig:entropy_2d} shows the radial entropy distribution
obtained in the 2D simulation (red and blue points) and the 1D
simulation (green line), at 50~ms, 100~ms, 200~ms, and 300~ms after
bounce. In general, the entropy spreads over a wider range in 2D than
in 1D, because of the convective activity between the shock and the
gain radius (i.e., the region where neutrino heating exceeds cooling).
The higher entropy in 2D is realized by more efficient neutrino
heating, in comparison to 1D \citep{hera94,burr95,jank96,murp08}.
Note that at late times, shown in the bottom panels of Figure
\ref{fig:entropy_2d}, once maximum entropy ($s\sim13~k_B$ per baryon
in the bottom left panel and 16 in the bottom right panel) is
obtained, matter at small radii ($\sim$ 100 km) does not gain
additional entropy/energy as it floats to larger radii. This is
related to the convection which drives matter from the gain radius,
where the heating is maximum, to larger radii, where the heating is
less efficient. Thus, the specific entropy is limited to the value at
the gain radius. This is illustrated in Figure~\ref{fig:entropy_2d},
where red points represent matter with positive radial velocity (i.e.,
escaping) that have high entropy, while blue points (accreting
material) have low entropy. For comparison, we show the weak
equilibrium entropy for the 1D simulation (green dotted line), that
would be achieved by infinitely-long exposure of matter to the
prevailing neutrino abundances and spectra. We construct this line by
keeping hydrodynamic quantities unchanged and solve the neutrino
emission and captures until equilibrium is almost achieved (see also
\citealt{lieb04}). The region where the entropy is below this line has
the potential to be heated by neutrino radiation up to the equilibrium
entropy, while the region above this line is dominated by neutrino
cooling. In 1D the entropy beyond the gain radius (corresponding to
the radius of the maximum entropy, $r\sim$ 100 km) decreases with
increasing radius, while the equilibrium entropy increases
continuously with radius. The difference between the obtained entropy
in the simulations and the equilibrium entropy is caused by the
insufficient time for the heating, i.e., too short advection timescale
compared to the neutrino heating timescale (see the discussion in
\citealt{bura06} for instance). Thus, the region above the gain radius
has room to absorb more neutrinos in multi-dimensional
simulations. The convective motion continuously exchanges equilibrated
matter from the gain radius by below-equilibrium matter from layers
that are too distant for efficient neutrino heating. This exchange of
matter by fluid instabilities enables continued neutrino heating at
the gain radius in combination with the steady distribution of the
heated matter to the outer layers \citep{hera94}.  Moreover, the
convective motion is expected to prolong the advection timescale and
lead to more effective neutrino heating. In summary, the higher
entropy obtained in 2D simulation is due to the presence of convection
and SASI in the gain region. However, even in 2D the advection
timescale is finite so that the equilibrium entropy is not realized.
Part of the material just below the gain radius maintains a higher
entropy than the equilibrium value because of the overshooting of high
entropy matter by convective motion. This phenomenon disappears deeper
inside the gain region ($r\lesssim 50$ km) because the weak
interaction rates at high density become large so that the entropy
coincides with the value determined by weak and thermal equilibrium.

\begin{figure*}[tbp]
\centering
\subfigure[LS180]{\includegraphics[width=0.48\textwidth]{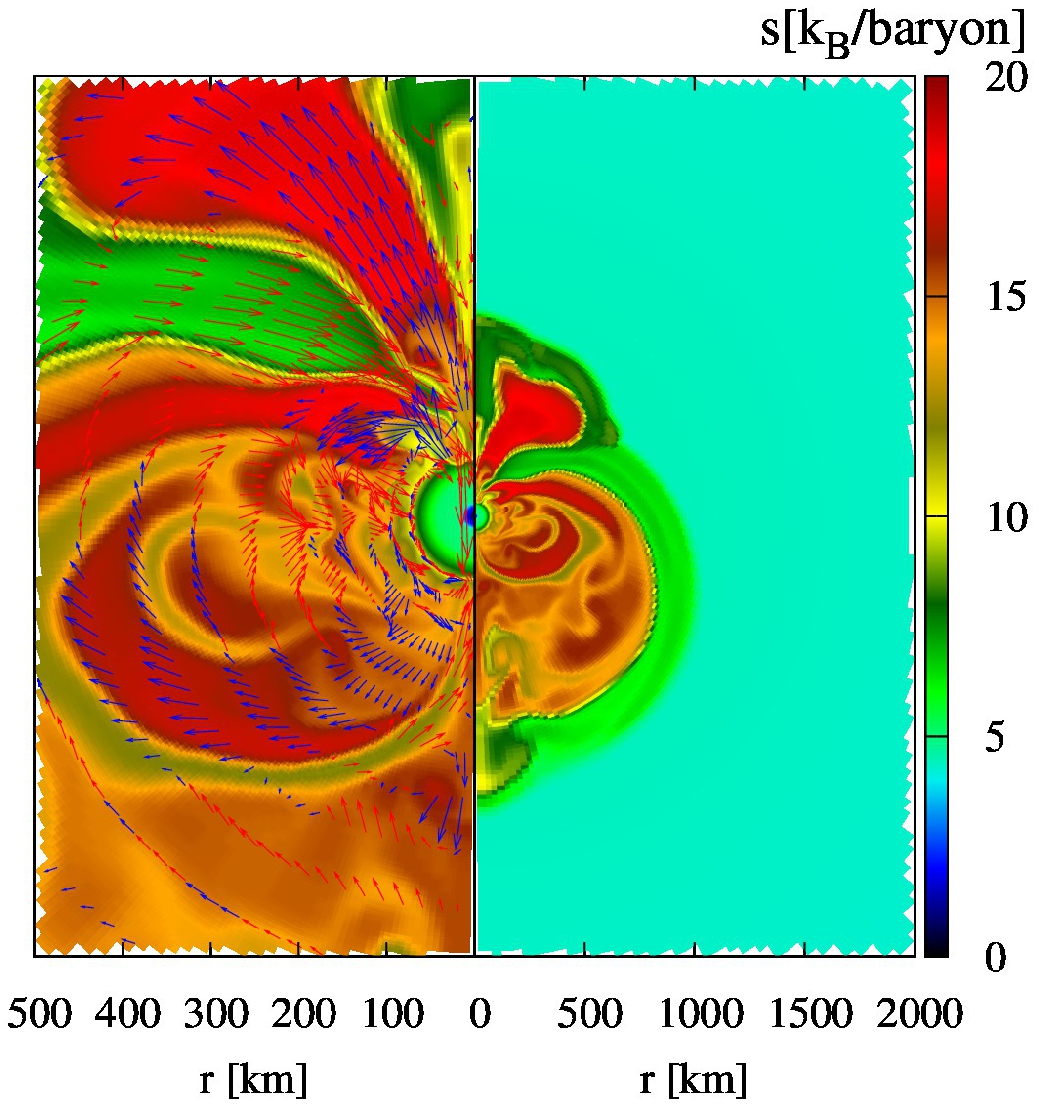}\label{fig:2d-entropy-ls}}
\hfill
\subfigure[SHEN]{\includegraphics[width=0.48\textwidth]{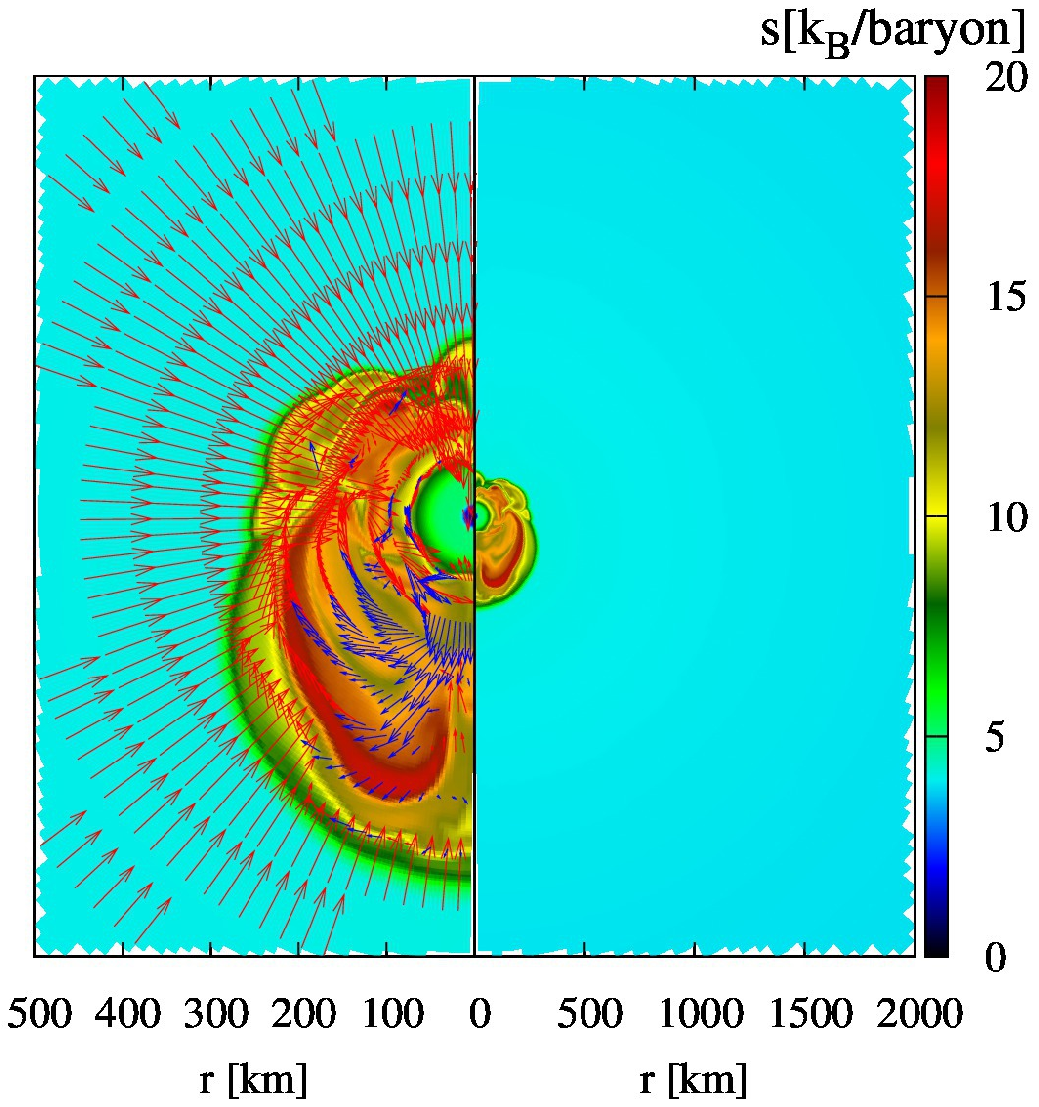}\label{fig:2d-entropy-shen}}
\caption{Snapshot at 400~ms post bounce of the entropy distributions
  comparing LS180 and SHEN.  Red and blue arrows show the in- and
  outflowing velocity fields, respectively. The left panel shows the
  inner part while the right panel is a wider view. The shock wave is
  strongly deformed due to SASI activity and turbulent matter motion
  induced by convection.  In the left panel we can see a cold downflow
  of matter that comes from the northern hemisphere, which activates a
  pressure wave in the southern hemisphere (see text for detail).
  This feature originates from the triple point of the shock wave that
  is shown in the northern hemisphere of the right panel.}
\label{fug:2d-entropy}
\end{figure*}

In Figure~\ref{fig:2d-entropy-ls} we show the time snapshot of the
entropy distribution at 400~ms after bounce.  The left panel shows the
central part and the right panel is a wider view.  In the left panel,
red arrows indicate infalling matter (negative radial velocity) and
blue arrows outstreaming matter (positive radial velocity). An
important structure becomes visible: low-entropy infalling matter
(``downflow'', see \citealt{hera92,burr95,jank96}) is accreting onto
the central PNS as indicated by the green area in
Figure~\ref{fig:2d-entropy-ls} (left panel). On the right-hand-side
panel of Figure \ref{fig:2d-entropy-ls}, a triple shock point is seen
to be located in the northern hemisphere near the position of (600 km,
200km). If one now compares with the left-hand-side panel of Fig 9a,
one recognizes that the cold downflow through the triple point advects
directly to the surface of the PNS. Hence, the cold downflow stems
from large distances, between 500 and 1000~km, and reaches down to the
PNS surface.  It exists as a quasi-stationary structure and supplies
energy and momentum to the PNS in an asymmetric way.  The momentum is
reflected at the surface of the PNS and becomes visible as a pressure
wave in Figure~\ref{fig:prompt}, which pushes the shock wave to larger
radii and increases the amplitude of the SASI by the so-called
``advective-acoustic cycle''~\citep{fogl07}. This feedback cycle
stands at the origin of the anisotropic shock evolution between north
and south poles which significantly promotes the shock expansion
reaching larger radii at the south pole.
 
\begin{figure}[tbp]
\centering
\includegraphics[width=0.48\textwidth]{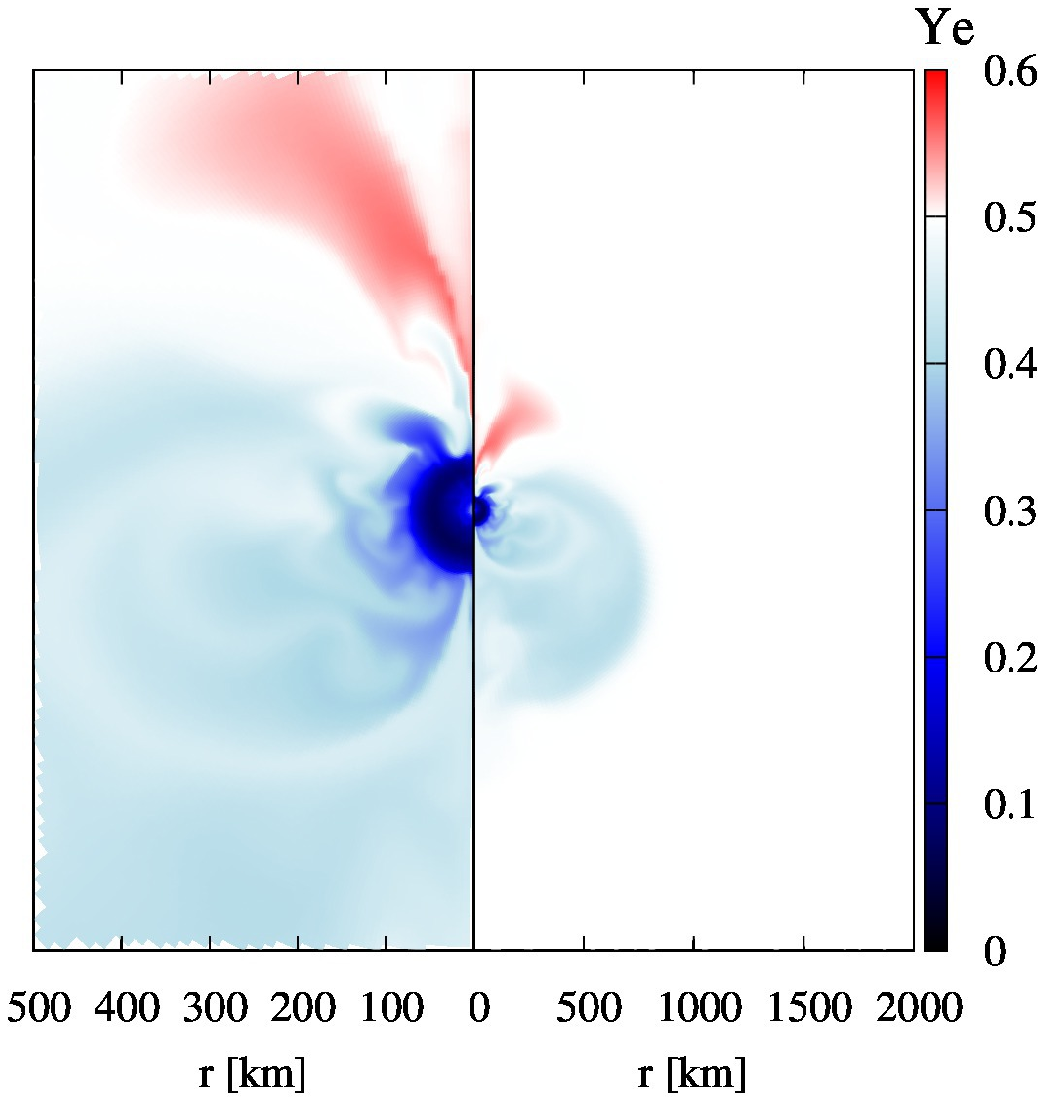}
\caption{$Y_e$ distribution for the simulation using LS180 at 400~ms
  post bounce.  Escaping material exhibits proton-rich conditions with
  $Y_e>0.5$, while infalling material becomes neutron rich due to
  electron captures.}
\label{fig:2d_ye}
\end{figure}

The IDSA neutrino transport scheme employed in this study permits an
estimate of the proton-to-baryon ratio of the possibly ejected
material from the PNS surface. In Figure~\ref{fig:2d_ye} we show the
$Y_e$ distribution at 400~ms post bounce. Infalling material at large
radii, dominated by heavy nuclei, starts with the progenitor value of
$Y_e\simeq 0.5$.  The electron fraction then decreases continuously
due to electron captures as the material is compressed while accreting
onto the PNS surface, where $Y_e$ finally reaches values as low as
$\simeq$ 0.05--0.2. However, the high-entropy region, which
corresponds to expanding matter, experiences a continuous flux of
neutrinos that stream off the PNS surface. There, in the presence of
similar electron neutrino and antineutrino luminosities, and small
average energy differences below 4~MeV between $\nu_e$ and
$\bar\nu_e$, matter becomes proton rich. Similar results have been
reported in previous studies of massive star explosions that employ
sophisticated neutrino transport \cite[e.g.,][]
{lieb01,bura06a,froh06,prue06,wana06,fisc10}.

In contrast to spherically symmetric models, multi-dimensional models
permit mass outflow alongside accreting matter and hence, e.g., mixing
with the surrounding material. Mixing opens interesting aspects for
the chemical evolution of explosion models which cannot be obtained
using 1D models. This has been discussed at the example of the 8.8
$M_\odot$ O-Ne-Mg-core explosion model in \citet{wana11} by comparing
1D and 2D results. Neutron-rich conditions were obtained in both cases
due to the extremely fast initial expansion of the ejecta. However,
while only slightly neutron-rich conditions could be obtained in 1D,
mushroom-like and more neutron-rich pockets developed in 2D.

\subsection{Comparison of 2D models with different EOS}

\begin{figure}[tbp]
\includegraphics[width=0.48\textwidth]{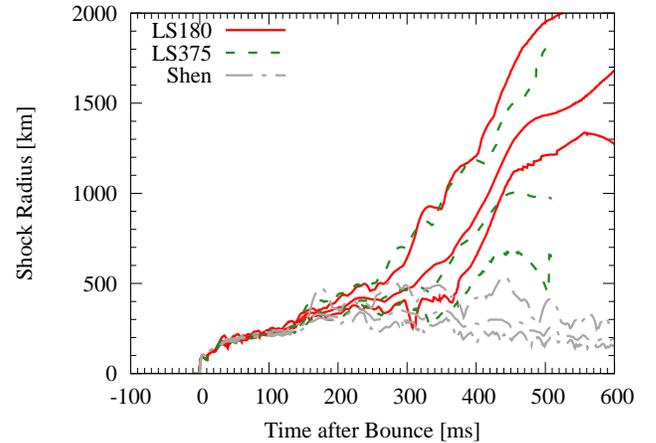}
\caption{Shock trajectories for the 2D simulations using the different
  EOS LS180 (red-solid lines), LS375 (green-dotted lines), and SHEN
  (gray dot-dashed lines). We show three lines for each model, which
  correspond to the maximum shock radius, the angular averaged one,
  and the minimum radius, from top to bottom.}
\label{fig:shock_2D}
\end{figure}

Here, we compare the 2D results for the 15 $M_\odot$ progenitor using
the different EOS, i.e., LS180, LS375, and SHEN.
Figure~\ref{fig:shock_2D} shows the shock-radius evolution for each
model.  Each of which has three lines, corresponding to maximum,
angular averaged and minimum shock radius (from top to bottom).  Note
that the minimum shock radius often coincides with the position of the
triple point of the shock, which is the starting point of the cold
downflow (see Figure \ref{fig:2d-entropy-ls}).  In
Figure~\ref{fig:shock_2D}, we can see that LS180 and LS375 (red solid
lines and green dotted lines) show similar trajectories.  There is a
gradual expansion phase during the post-bounce time
$t_\mathrm{pb}\lesssim 200$--300~ms, where the maximum shock radii
grow to about 500~km. While afterwards, they grow more rapidly in both
models and reach about 2000~km for LS180 and 1800~km for LS375 at
500~ms post bounce. These two models show no phase of shock
retraction. For the 2D simulation using SHEN, on the other hand, the
shock wave does not enter this second phase of fast shock expansion
even at late times (see Figure~\ref{fig:shock_2D}).

Before analyzing the the origin of this observed difference in the
shock evolution between SHEN and LS180 further, a comment regarding
the likelihood to find explosions in our models seems in order: At the
time the 2D simulations are stopped, the ``diagnosing explosion
energy'', as defined in \citealt{suwa10}, achieves $\sim 10^{50}$~erg
in both LS EOS cases, while explosions for the 2D simulations using
SHEN are unlikely to occur. LS180 has been applied in axially
symmetric simulations of the same 15 $M_\odot$ progenitor in
\citet{mare09}, where the average shock radius was found to oscillate
around 200~km for several 100~ms until it eventually starts to expand
at about 600~ms post bounce, i.e. much later than in our
simulations. The fast and continuous shock expansion to increasingly
larger radii for our LS180 simulation is likely due to the omission of
the continuous energy loss by the emission of heavy lepton neutrinos,
and possibly the neglect of general relativistic effects in the
gravitational potential.  Coming back to the discussion of the
differences between LS180 and SHEN, we will analyze in the following
the two different aspects of (a) neutrino heating/cooling, and (b)
subsequent convective activity.

\begin{figure*}
\subfigure[$\nu_e$ average energy]{\includegraphics[width=0.48\textwidth]{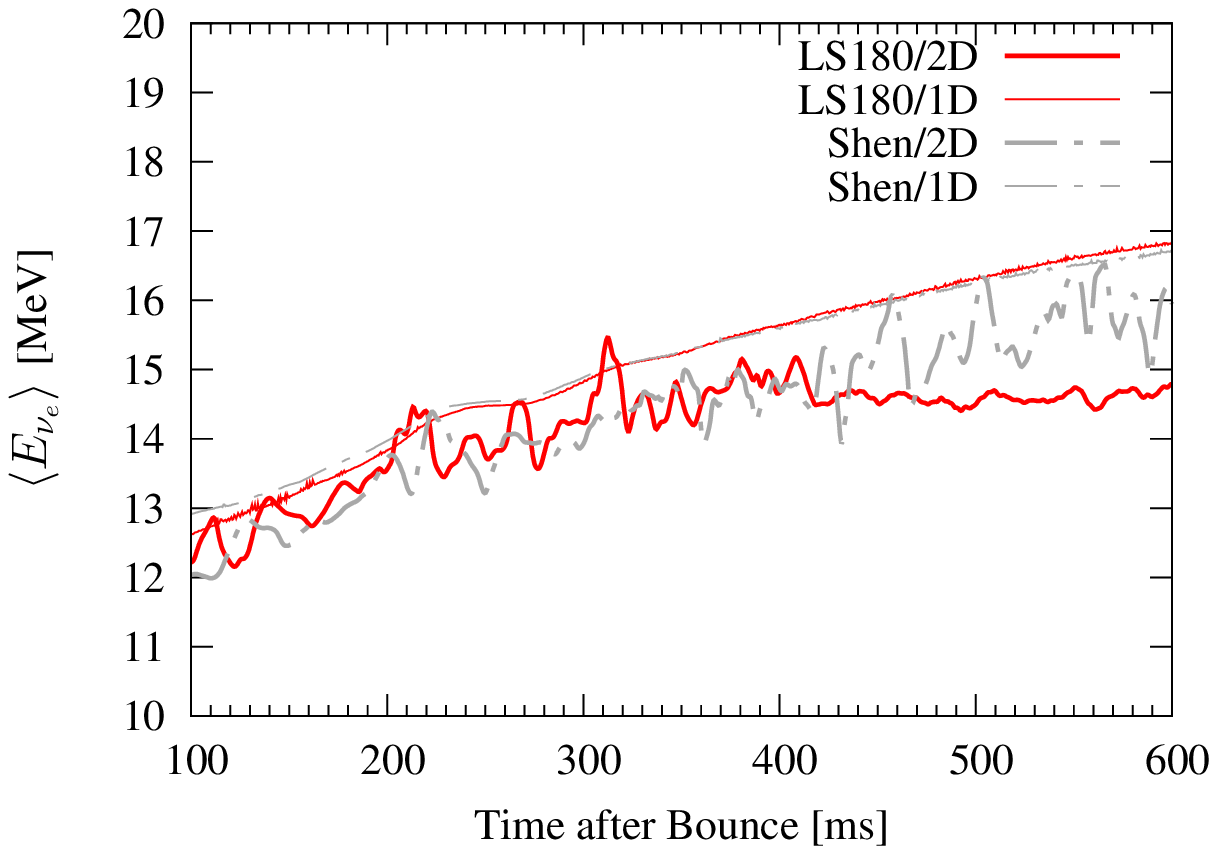}
\label{fig:eve-2d}}
\subfigure[$\bar\nu_e$ average energy]{\includegraphics[width=0.48\textwidth]{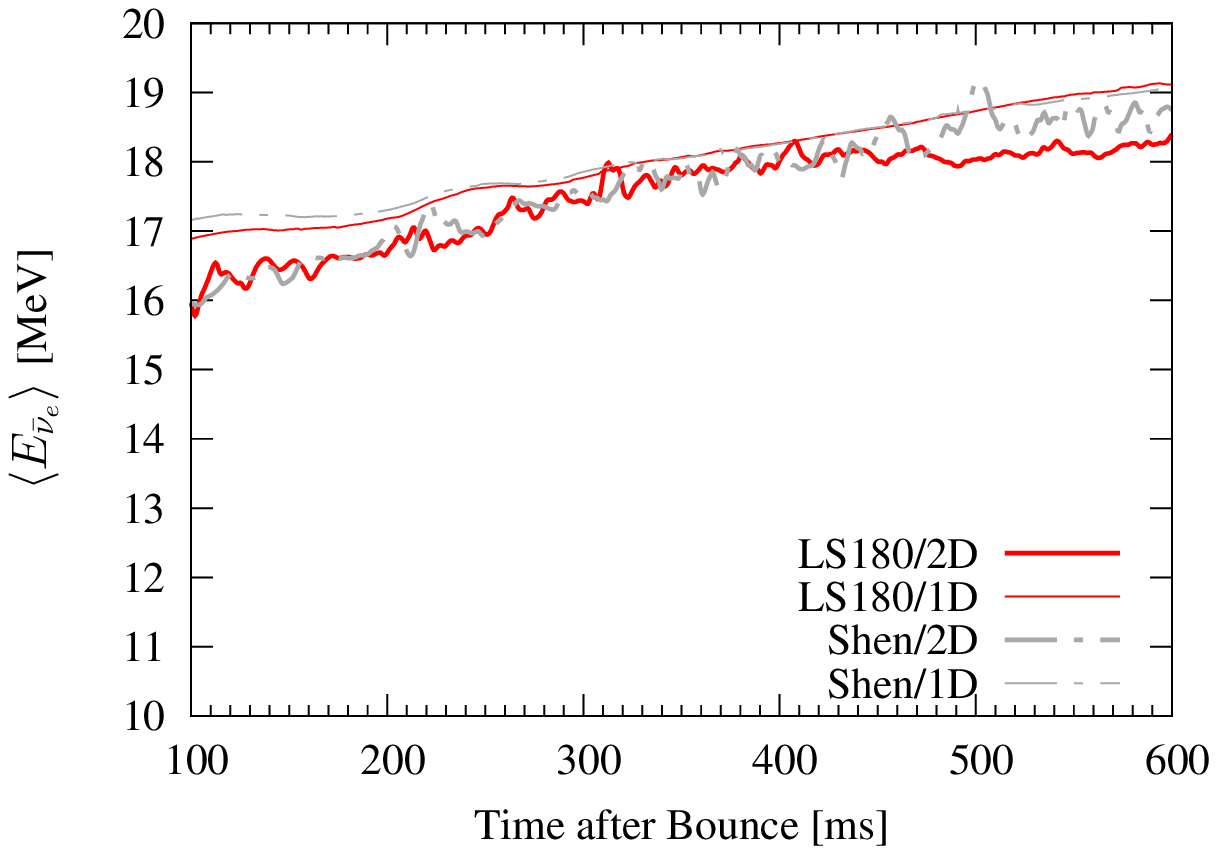}
\label{fig:eveb-2d}}
\hfill \subfigure[$\nu_e$ luminosity]{\includegraphics[width=0.48\textwidth]{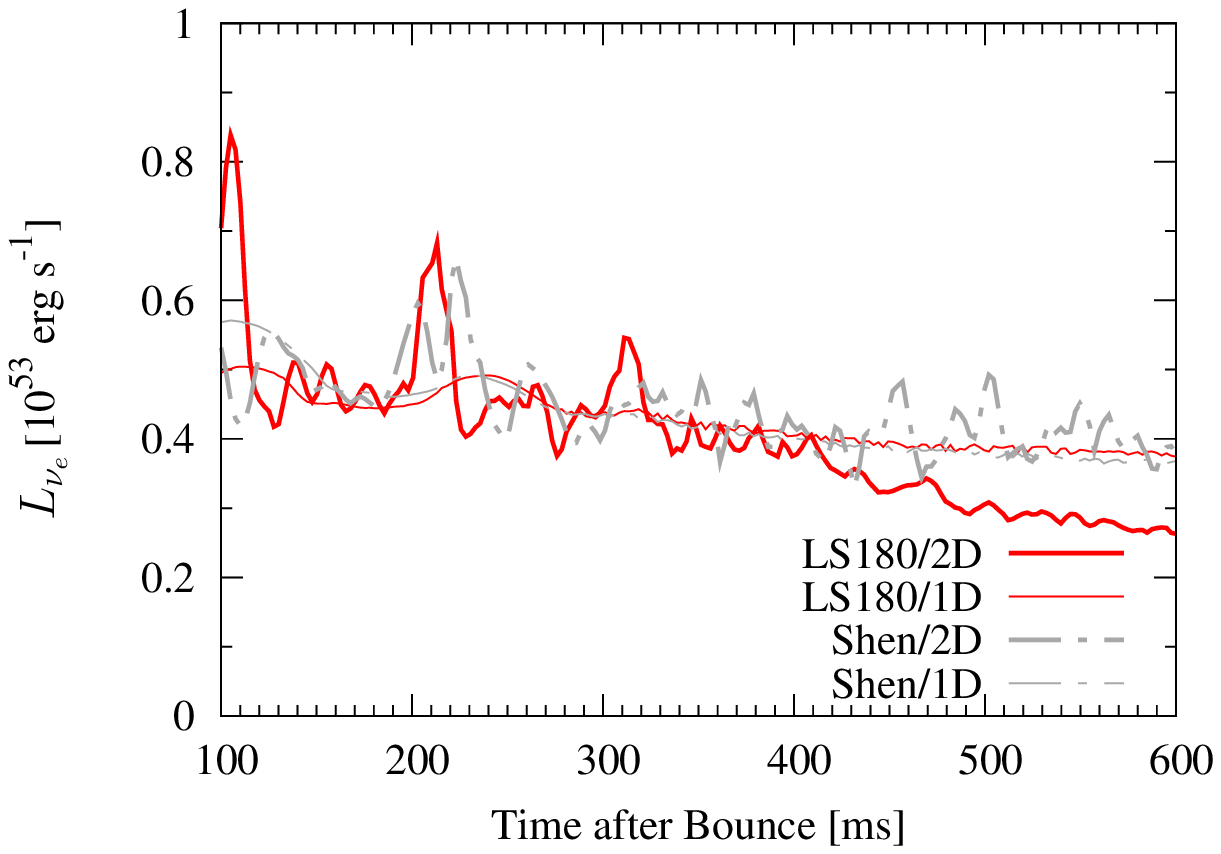}
\label{fig:Lve-2d}}
\hfill \subfigure[$\bar\nu_e$ luminosity]{\includegraphics[width=0.48\textwidth]{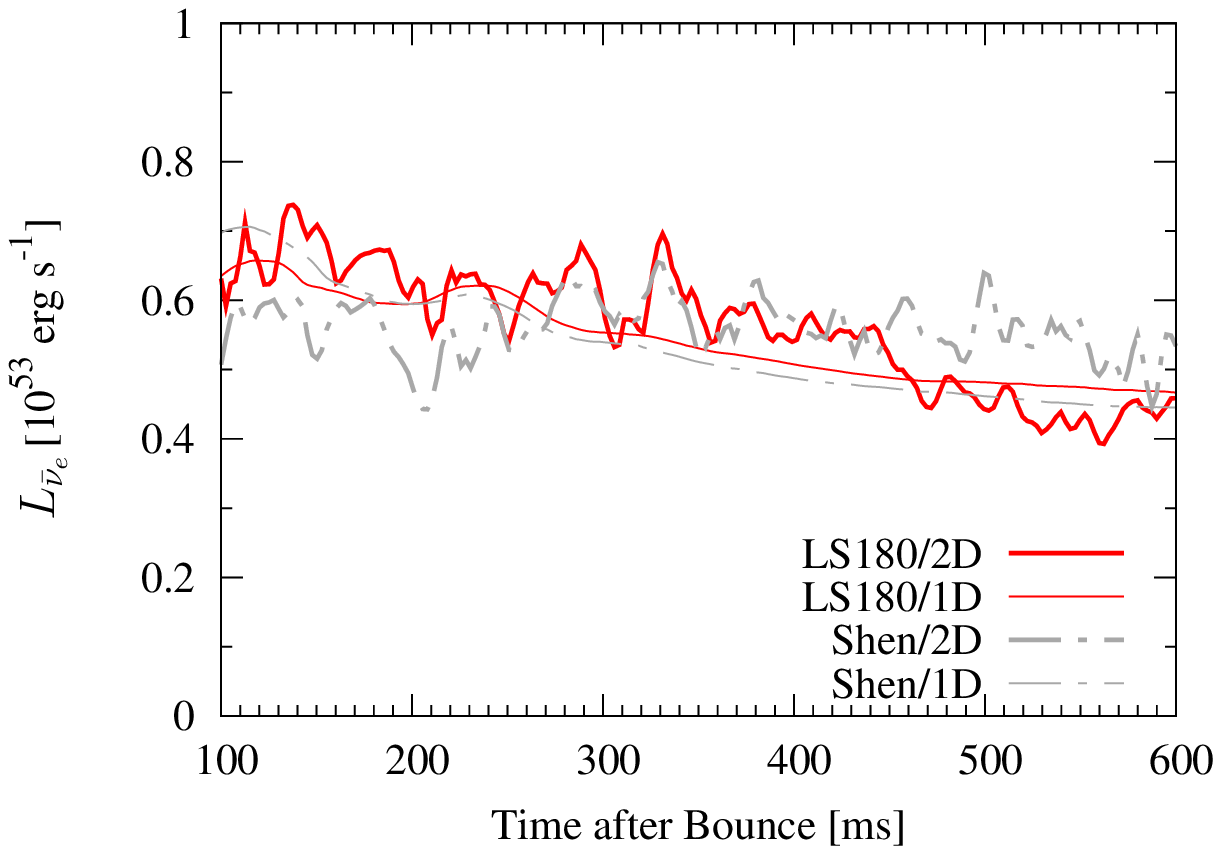}
\label{fig:Lveb-2d}}\\
\caption{Post-bounce evolution of neutrino mean energies and
  luminosities, for the 2D simulation using LS180 and SHEN sampled at
  the equator (thick lines). For comparison, we also show results from
  the 1D simulation using the same EOS (thin lines).}
\label{fig:nu-2d}
\end{figure*}

Figure~\ref{fig:nu-2d} illustrates the evolution of the neutrino
luminosities and average energies in 1D (thin lines) and 2D (thick
lines), comparing LS180 (red solid line) and SHEN (grey dot-dashed
line), for electron neutrinos (left panel) and antineutrinos (right
panel), respectively.  Note the rapid variations in the luminosities
on a millisecond timescale, which are due to the convective activity
in the vicinity of the PNS (e.g., \citealt{mare09b}). At first, the
luminosities (sampled at the equator) are in qualitative agreement
with the results obtained in spherical symmetry.  The mean neutrino
energies in the 2D models are slightly, but systematically, smaller
than in the 1D models. When the standing shock starts to propagate
faster to increasingly larger radii in 2D, i.e.  at about 350~ms post
bounce for LS180, the $\nu_e$ and $\bar\nu_e$ energies cease to rise
and stay nearly constant (see Figs.~\ref{fig:eve-2d} and
\ref{fig:eveb-2d}).  Furthermore, the luminosities decrease rapidly
and deviate from the 1D results (see Figs.~\ref{fig:Lve-2d} and
~\ref{fig:Lveb-2d}).  This phenomenon is more prominent for $\nu_e$,
whose neutrinosphere is located at lower densities than those of
$\bar\nu_e$.

%

The LS180 and SHEN EOS lead to different central configurations during
the post-bounce evolution, as a result of the different PNS
contraction behaviors.  In comparison to the stiffer SHEN EOS, the
soft LS180 EOS results in a more compact PNS with higher central
density and a steeper density gradient at its surface.  The impact of
different PNS structures on the properties of emitted neutrinos has
been discussed in many places of the literature \citep[see,
  e.g.,][]{hemp12}. A direct consequence of the different PNS
structures is the different evolution of the neutrino luminosities
(see Figs.~\ref{fig:Lve-2d} and \ref{fig:Lveb-2d}). In 1D, the
neutrino luminosities (thin lines) become slightly smaller for SHEN
compared to LS180. However, the difference between SHEN and LS180 is
much smaller than what has been found in \citet{sumi05, fisc09}. The
reason might be that our PNS is less compact due to the neglect of
heavy-lepton neutrinos and general relativistic gravity. As the shock
for LS180 in the 2D simulations starts to expand continuously after
about 350~ms post bounce, we will focus on post-bounce times before
that moment to search for the origin of more efficient neutrino
heating in LS180 compared to SHEN.  It can be seen in
Figs.~\ref{fig:Lve-2d} and \ref{fig:Lveb-2d} that the $\nu_e$
luminosities are rather similar for LS180 and SHEN, while the
$\bar\nu_e$ luminosities differ considerably during the first 250~ms
post bounce.  We explain these differences by the fact that the
$\bar\nu_e$ neutrinospheres are located at higher density where the
EOS differences between LS180 and SHEN are larger than at the lower
density where the $\nu_e$ neutrinospheres are located.  The sizeable
difference in the $\bar\nu_e$ luminosity leads to less efficient
heating for SHEN, and consequently to a lower entropy per baryon in
the gain region than in the simulations based for LS180 (see
Figure~\ref{fig:2d-entropy-shen}). How the luminosity decays at later
times, as observed here after the onset of the explosion in the
simulations using LS180, and as predicted by recent spherically
symmetric explosion models, needs to be elaborated by long-term
multi-dimensional simulations that cover several seconds after the
onset of explosion. This is yet computationally very expensive and
beyond the scope of the present study.

\begin{figure*}[tbp]
\subfigure[Integrated kinetic energy of the lateral direction.]
{\includegraphics[width=0.48\textwidth]{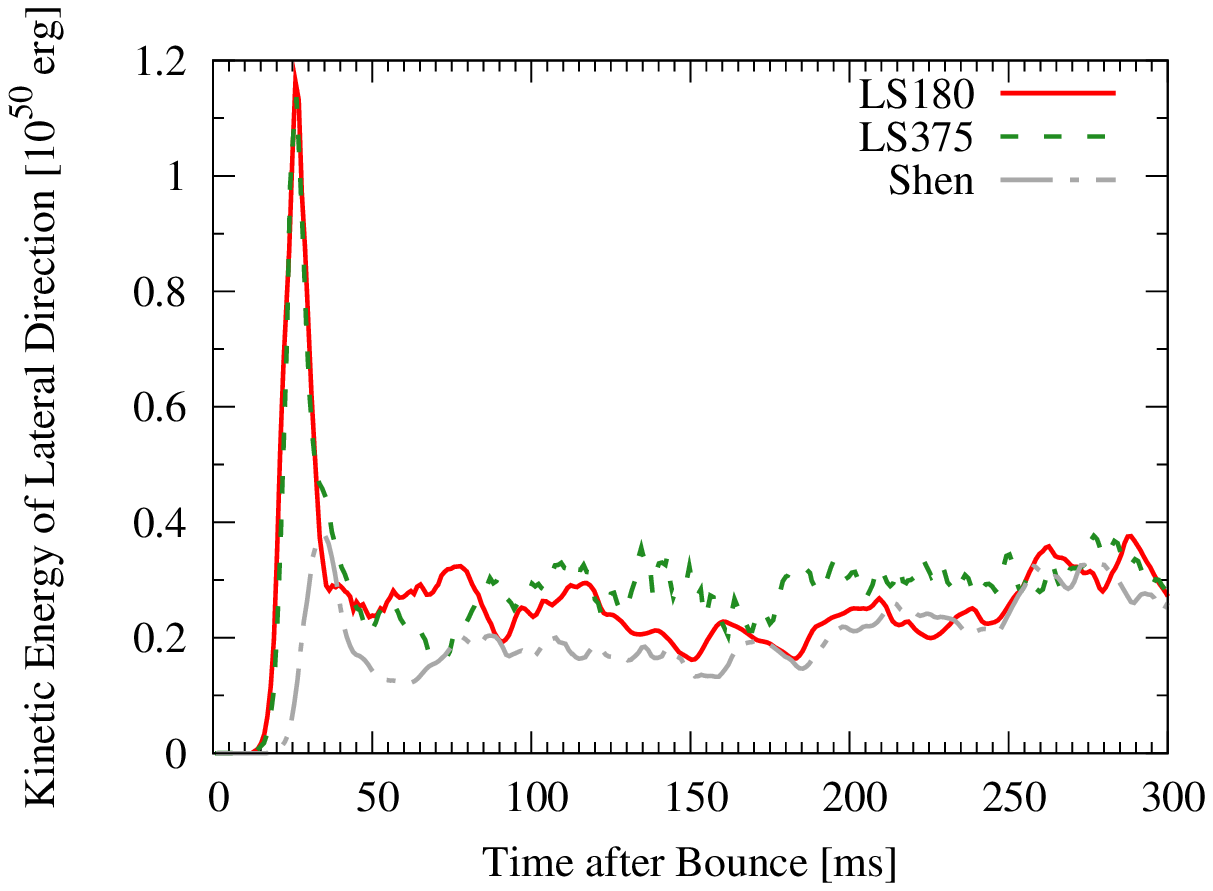}}
\hfil
\subfigure[Distribution of the root-mean-square of the radial velocity
  in a space-time diagram. The top half and bottom half correspond to
  LS180 and SHEN, respectively.]
{\includegraphics[width=0.51\textwidth]{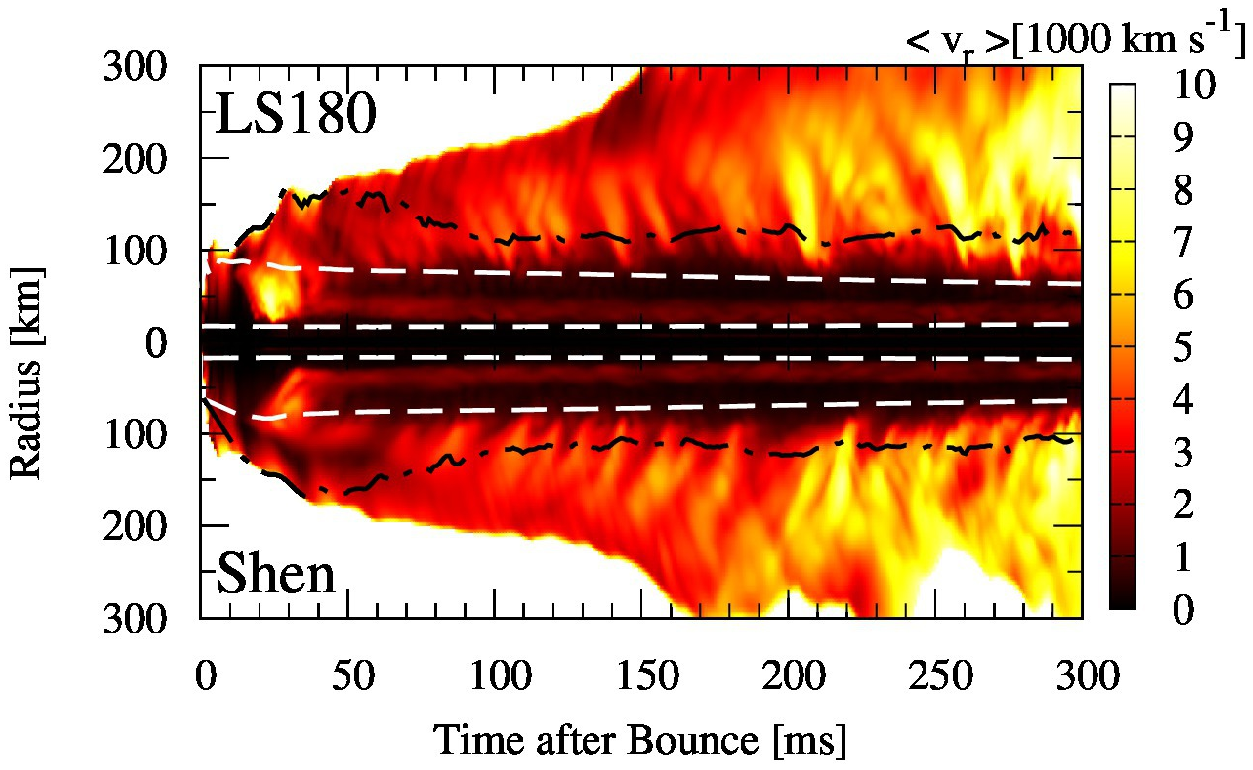}}
\caption{Post bounce evolution of kinetic energy and root-mean-square
  of the radial velocity field as indicators for convective
  activity. In graph~(a), both, LS180 (red solid line) and LS375
  (green dashed line), exhibit a strong activity, while the activity
  for SHEN (gray dash-dotted line) is weaker by a factor of $\sim 3$
  during the first 50~ms after bounce. After that, prompt convection
  vanishes and LS180 and LS375 maintain slightly larger kinetic energy
  than SHEN. In graph~(b), the colored region is the post-shock
  layer. The black dash-dotted lines indicate the position of the gain
  radius and the white dashed lines are the radii of
  $\bar\rho=10^{11}$ (outer) and $10^{13}$~g~cm$^{-3}$ (inner), where
  $\bar\rho$ is the angle averaged density. The advection-mediated
  convective motion (see text) is apparent for
  $10^{11}$~g~cm$^{-3}\lesssim \bar\rho\lesssim 10^{13}$~g~cm$^{-3}$
  and $t_\mathrm{pb}\lesssim 50$~ms for LS180. After such convective
  motions vanish, weak continued PNS activity is visible above
  $\bar\rho=10^{13}$~g~cm$^{-3}$.}
\label{fig:prompt}
\end{figure*}

In the following paragraph, we will return to aspect (b), which is a
comparison of convective activities between our 2D models using SHEN
and LS EOS. The left panel of Figure~\ref{fig:prompt} shows the
post-bounce evolution of the kinetic energy of the lateral ($\theta$)
direction, $\int \rho v_\theta^2/2~dV$, where $V$ is the entire
simulation domain. This quantity is zero in spherical symmetry such
that non-vanishing values imply a deviation from sphericity. Models
based on LS180 and LS375 exhibit a significantly larger amplitude than
SHEN for $t_\mathrm{pb}\lesssim 50$ ms. This is the so-called prompt
convection, which is triggered by the negative entropy gradient left
behind the prompt shock as it weakens by dissociation of nuclei and
eventually comes to a complete stall by the emission of the
deleptonization burst.  In the right panel of Figure~\ref{fig:prompt},
we show the radial profiles of the root-mean-square of the radial
velocity, $(\int v_r^2 \sin\theta d\theta)^{1/2}$. For LS180 (upper
part), convective motion is apparent between 30--100~km during the
first 50~ms after bounce, which is seen as a yellowish island in the
plot. This shows that the convective flows produced by the fluid
instabilities advect down to the inner surface of the PNS (white
dashed line). The occurrence of this advection-mediated convective
motion has been already mentioned in \citet{bura06,dessart06}.  For
SHEN (lower panel of Figure~\ref{fig:prompt}(b)), such a behavior is
only weakly visible due to the smaller volume of the convectively
unstable region.  Note that after the convection has ceased, both,
LS180 and LS375, still have larger kinetic energy than SHEN.

In order to analyze the evolution of the shock wave, we conduct a mode
analysis as follows: The deformation of the shock surface is
decomposed into spherical harmonic components
\begin{equation}
R_s(\theta)=\sum_{\ell=0}^\infty a_\ell\sqrt{\frac{2\ell+1}{4\pi}}P_\ell(\cos\theta).
\end{equation}
Note that due to axisymmetry, only $m=0$ harmonics are relevant. The
coefficients, $a_\ell$, can be calculated by the orthogonality of the
Legendre polynomials,
\begin{equation}
a_\ell=\frac{2\ell+1}{2}\int^{1}_{-1}R_s(\theta)P_\ell(\cos\theta)d\cos\theta.
\end{equation}
The position of the shock surface, $R_s(\theta)$, is estimated from
the isentropic surface of $s=6 k_B$ per baryon.

\begin{figure}[tbp]
\includegraphics[width=0.48\textwidth]{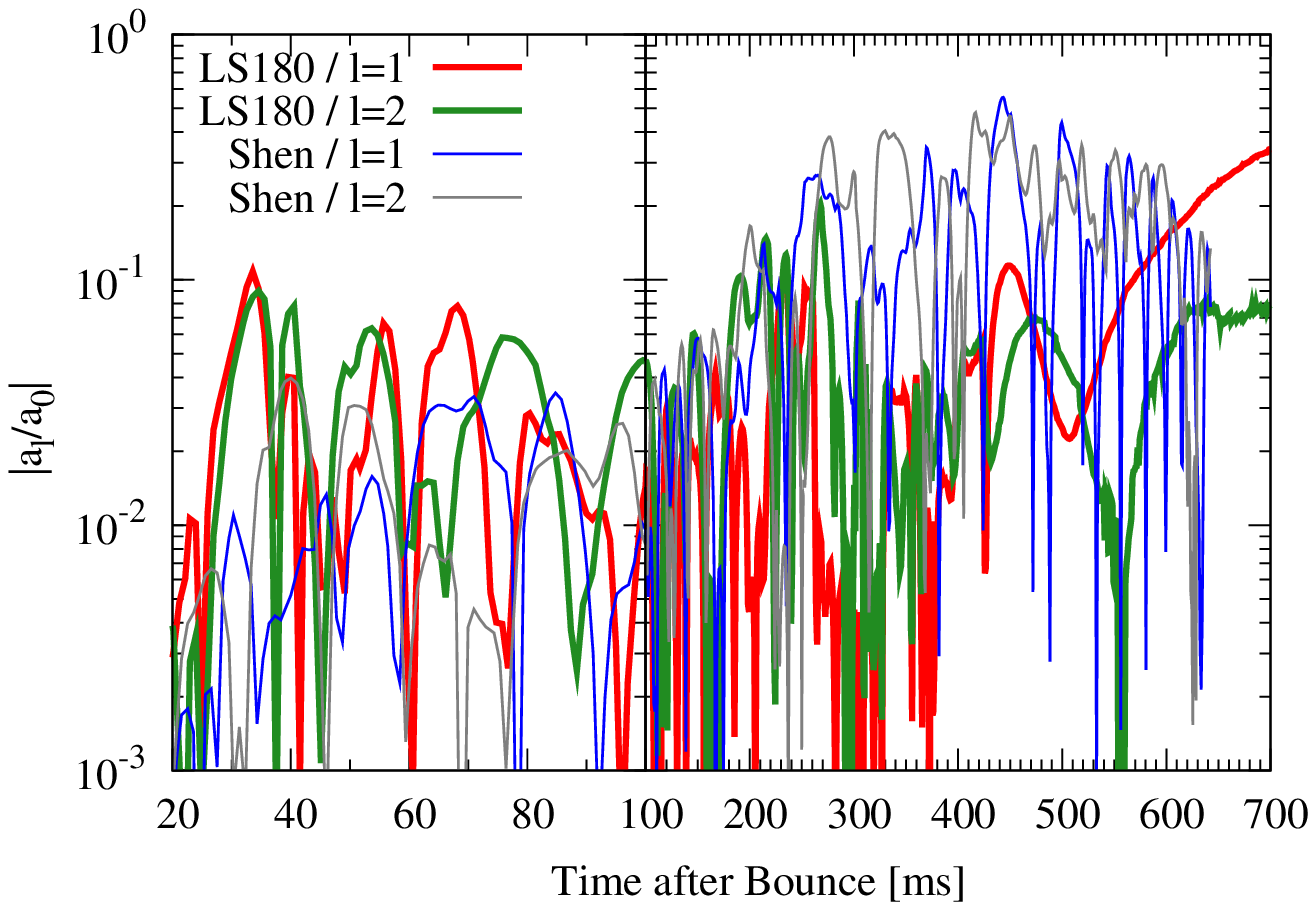}
\caption{Post-bounce evolution of the amplitude $\ell=1$ and 2 SASI
  modes, comparing LS180 (thick lines) and SHEN (thin lines). During
  the early phase ($t_\mathrm{pb}\lesssim$ 100 ms) LS180 has a larger
  amplitude due to the strong prompt-convection activity. Later, LS180
  exhibits a smaller amplitude due to the continuously expanding shock
  wave to larger radii, while SHEN has still an increasing
  amplitude. However, also for SHEN, it saturates and begins to
  decrease at $t_\mathrm{pb}\gtrsim$ 400~ms, indicating a failed
  explosion.}
\label{fig:sasi}
\end{figure}

Figure~\ref{fig:sasi} shows the post-bounce evolution of the low mode
($\ell=1$ and 2) for LS180 and SHEN. During the early phase
($t_\mathrm{pb}\lesssim$100 ms after bounce), LS180 has a
significantly larger amplitude than SHEN.  After that the amplitudes
are similar in the time window $100\lesssim t_\mathrm{pb}\lesssim 300$
ms. For LS180, the shock wave begins to expand more rapidly at
$t_\mathrm{pb}\sim 300$~ms, after which the amplitude is decreasing
continuously with time. For SHEN, the shock wave continues to
oscillate at smaller radii and does not expand. This in turn results
in an increasing amplitude.  At $t_\mathrm{pb}\gtrsim 500$ ms, LS180
shows a dominant $\ell=1$ mode (red line) due to the expansion of the
shock, while SHEN shows a decreasing amplitude because the retracting
shock wave is approaching a spherical shape.  This corroborates the
failed explosion for SHEN. Note that the mode analysis of the shock
radius, albeit very helpful to extract the information of SASI modes,
serves as a reliable indicator of the explosion only if it is combined
with other criteria such as the convention timescale argument
regarding the heating/advection timescale and the antesonic condition
that we will discuss later in this section.

\begin{figure*}[htbp]
\centering
\includegraphics[width=1.\textwidth]{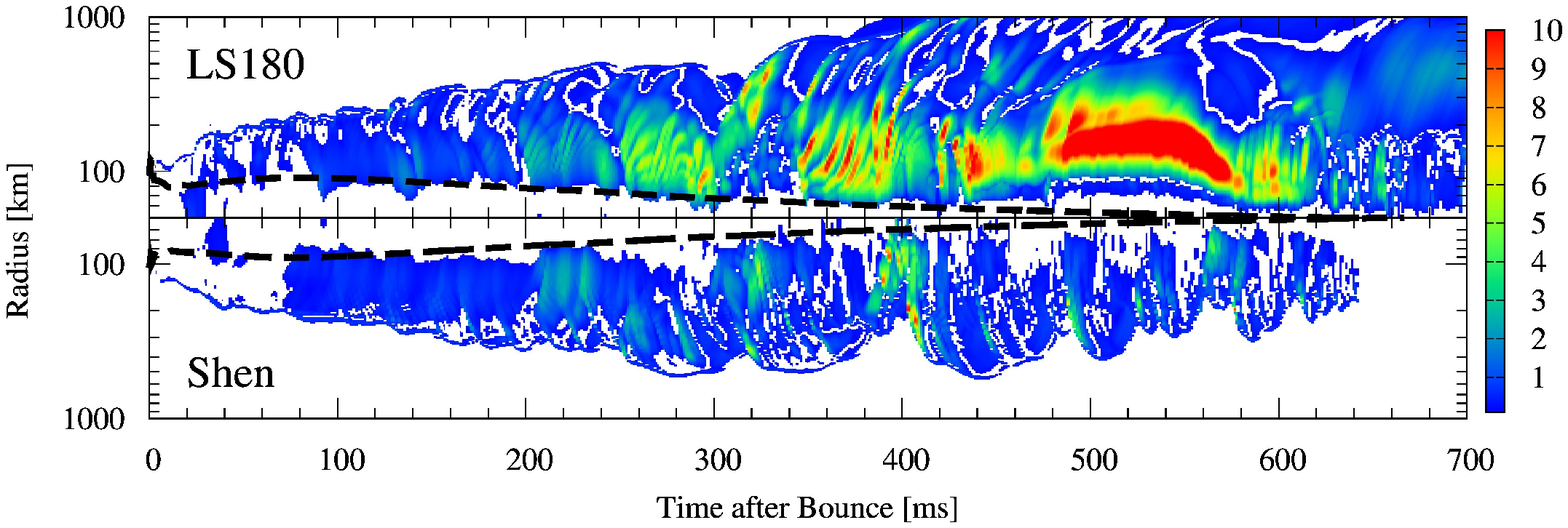}
\caption{ Post bounce evolution of the dispersion of $\mathcal{M}$
  (Eq. \ref{eq:dis_M}) for LS180 (top panel) and SHEN (bottom
  panel). The white color is set below the minimum value of 0.1. The
  thick dashed lines show the position of the neutrinosphere
  representing the surfaces of the PNS.}
\label{fig:p}
\end{figure*}

Next, we discuss the EOS dependence of the SASI activity
(acoustic-vorticity cycle). We show the {\it pressure} deviation for
LS180 and SHEN in Figure~\ref{fig:p}. It is determined as follows:
\begin{eqnarray}
\frac{\left\{\frac{1}{2}\int^\pi_0 \left[\mathcal{M}(r,\theta)-\overline{\mathcal{M}}(r)\right]^2\sin\theta d\theta\right\}^{1/2}}{\overline{\mathcal{M}}(r)},
\label{eq:dis_M}
\end{eqnarray}
where
\begin{eqnarray}
&&\mathcal{M}(r,\theta)\equiv\rho(r,\theta)v_r^2(r,\theta)+P(r,\theta),\\
&&\overline\mathcal{M}(r)\equiv\frac{1}{2}\int^\pi_0 \mathcal{M}(r,\theta)\sin\theta d\theta.
\end{eqnarray}
$\mathcal{M}$ is the total pressure, including the ram pressure of the
infalling material. It should assume the same value ahead and behind
the shock wave (corresponding to the momentum part of the
Rankine-Hugoniot equation). Therefore, $\mathcal{M}$ is a useful
quantity to investigate the shock expansion: If $\mathcal{M}$ is
greater behind the shock than ahead of the shock, the shock wave
propagates outward, and vice versa. Before 100~ms postbounce, the
dispersion of $\mathcal{M}$ is small ($\sim 0.1$) (especially for
SHEN; see Figure~\ref{fig:p}). Later, the pressure perturbation
increases in the shocked region and the shock wave is pushed outward.
For LS180, the pressure perturbation continues to grow after
$t_\mathrm{pb}\sim200$--300~ms and a strong pressure wave transfers
momentum to the shock. This phenomenon is absent in the simulations
using SHEN. As a consequence in the LS180 model, the pressure behind
the shock wave becomes greater than the ram pressure ahead of the
shock, so that the shock begins to propagate outward without receding.
Note that the pressure wave appears to be produced close to the PNS
surface (see the thick dashed lines in Figure~\ref{fig:p}), where the
vorticity is reflected at the steep density gradient (see also the
right panel of Figure \ref{fig:prompt}).

\begin{figure*}[tbp]
\subfigure[Accretion ratio determined by
  Eq.~(\ref{eq:acc_ratio}).]{\includegraphics[width=0.48\textwidth]{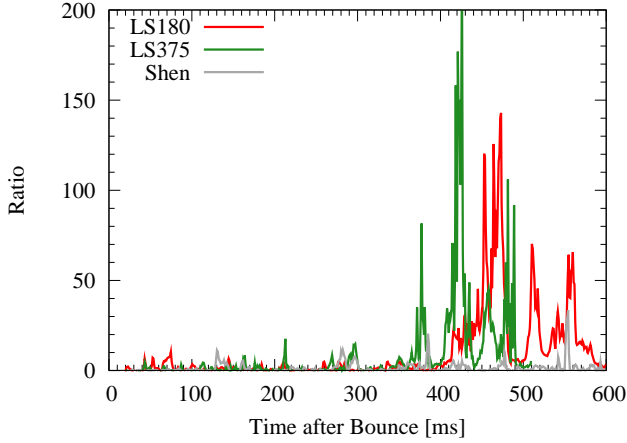}
\label{fig:kine}}
\hfill \subfigure[Mass-weighted average entropy per baryon in the gain region.]{\includegraphics[width=0.48\textwidth]{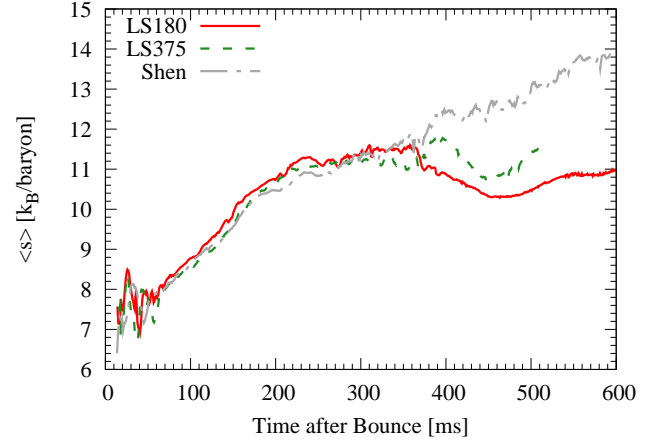}
\label{fig:save}}
\\
\subfigure[Mass enclosed in the gain region.]{\includegraphics[width=0.48\textwidth]{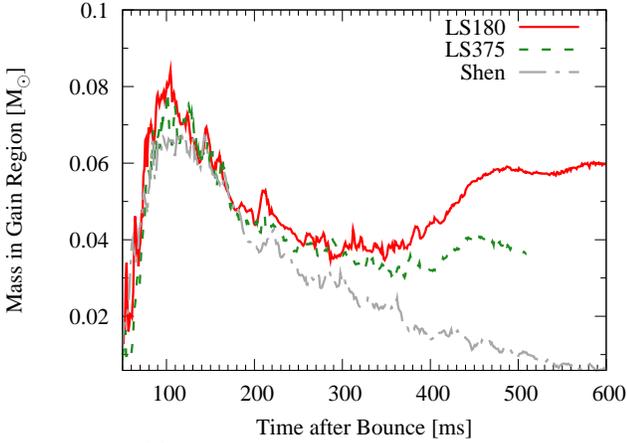}
\label{fig:gmass}}
\hfill 
\subfigure[Protoneutron star radii, determined at $\rho=10^{11}$~g~cm$^{-3}$.]{\includegraphics[width=0.48\textwidth]{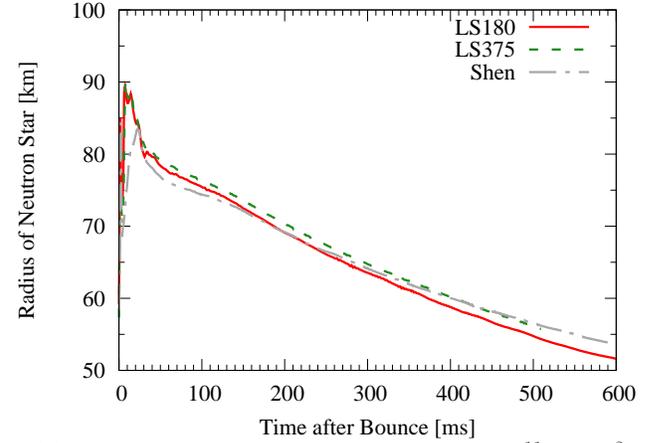}
\label{fig:nsrad}}
\caption{Post-bounce evolution of selected quantities. In graph~(a),
  the models using the LS EOS have a larger anisotropic downflow,
  i.e. the ratio becomes larger than 100, while SHEN lies between 0.1
  and 10 so that the downflow is rather spherical, slightly
  oscillating from north pole to south pole and vice versa on a
  timescale on the order of 100~ms.}
\end{figure*}

In order to characterize the spatial anisotropy of downflows for the
different EOS used, we estimate the following ratio, $\mathcal{R}$,
which is determined by
\begin{equation}
\mathcal{R}\equiv\frac{1}{2}\left(\frac{E^N_K}{E^S_K}+\frac{E^S_K}{E^N_K}\right)-1,
\label{eq:acc_ratio}
\end{equation}
where $E^N_K$ and $E^S_K$ are the kinetic energy accreted from the
northern $(N)$ and southern $(S)$ hemispheres, respectively. $E_K=\int
\rho v_r^2/2~dA$, where the integration extends over the surface of a
sphere with a radius of 100~km. In Figure~\ref{fig:kine}, we show the
post-bounce evolution of $\mathcal{R}$, comparing LS180, LS375, and
SHEN. It implies that LS180 and LS375 have greatly asymmetric mass
accretion after 300--400~ms post bounce, while for SHEN the
mass-accretion asymmetry is small.

In the following paragraphs, we investigate possible indicators of the
onset of explosion which have been previously explored in the
literature.  One of those is the mass-weighted average entropy in the
gain region, discussed as an indicator of the neutrino-heating
efficiency in \cite{murp08} and \cite{nord10}.  It has been suggested
that exploding models exhibit higher entropy in the gain region prior
to the onset of explosion, as a result of more efficient neutrino
heating, in comparison to non-exploding models of the same progenitor.
Figure~\ref{fig:save} shows the post-bounce evolution of the average
entropy, $\bracket{s}$, comparing LS180, LS375, and SHEN.  In
agreement with previous studies, we find that the optimistic
simulations, which show a continuous shock expansion to increasingly
large radii (LS180, LS375), have higher average entropy than the
less-optimistic one (SHEN), until the onset of explosion at about
350~ms for LS180.  Once the explosion has been launched, the
entropy-growth rate slows down and even decreases as a consequence of
a continuous accretion of cold matter from the upwind side.  Note that
the entropy per baryon will rise again at the moment when mass
accretion will turn into mass outflow, such that net neutrino heating
establishes at the surface of the PNS. A similar phenomenon has been
observed when increasing the dimensionality in simulations of
neutrino-driven supernova explosions, e.g., from axial symmetry to
three dimensions in parametric studies \citep{nord10,hank12}.  The
presence of convection and SASI exposes matter in the gain region to
net neutrino heating for a longer period. In our models based on
LS180, the stronger convective activity leads to more efficient
neutrino heating for matter in the gain region such that a higher
entropy per baryon is achieved than in the models based on SHEN.

Another important indicator to predict neutrino-driven explosions is
the mass enclosed in the gain region \citep[see, e.g.,][]{jank01}. In
Figure \ref{fig:gmass} we find that it is always smaller for SHEN than
for LS180 and LS375, because SHEN produces a less compact object with
lower central density and a less steep density gradient at the PNS
surface than LS. The mass enclosed inside the gain region reaches its
maximum at about 100~ms post bounce, about 0.065 $M_\odot$ for SHEN
and 0.085 $M_\odot$ for LS. This time corresponds to the time with the
maximum accretion rates (see the accretion luminosities in
Figure~\ref{fig:acc_lum}). After that, the mass enclosed inside the
gain region decreases. After about 150~ms post bounce the enclosed
mass in the gain region for SHEN falls significantly below the
corresponding values for LS. Note that otherwise the evolutions for
SHEN and LS are rather similar up to this moment, see
Figure~\ref{fig:p}, but about 100~ms later they differ
substantially. As discussed above, this is a consequence of more
efficient neutrino heating for LS due to a higher $\bar\nu_e$
luminosity on the one hand and a larger mass enclosed inside the gain
region on the other hand. For SHEN, the enclosed mass inside the gain
region drops below 0.02 $M_\odot$ between 200--300~ms after bounce,
while it turns around at a value of 0.04 $M_\odot$ for LS (see
Figure~\ref{fig:gmass}). As the shock starts to expand for LS only,
the gain-region mass differences between SHEN and LS diverge to an
extent that it is not meaningful to compare them at a later postbounce
time.  In summary, the different structures of the protoneutorn stars
for SHEN and LS lead to a smaller mass enclosed inside the gain region
and lower luminosities (mainly $\bar\nu_e$) for SHEN. Both effects
inhibit the efficiency of the neutrino heating in the models based on
SHEN, that do not lead to an explosion in our 2D models.

Additionally the less efficient heating for SHEN can also be related
to the absence of large-scale hydrodynamic instabilities.
Deformations of the standing-accretion shock assisted by buoyant
convections are significantly smaller for SHEN compared to LS180
(compare Figure~\ref{fig:2d-entropy-ls} and
\ref{fig:2d-entropy-shen}).  Moreover, we cannot find any large
down-streams of material for SHEN, which for LS180 supply continuously
energy to the central regions in an asymmetric way.

Because the structure of the PNS now appeared several times to stand
at the origin between runs using LS180 and SHEN, we wonder whether the
radius of the PNS might serve as a reliable explosion indicator. The
post-bounce evolution of the PNS radius is shown in
Figure~\ref{fig:nsrad}. After the rapid initial rise up to about 90~km
for the models using LS180 and LS375, between 10--20~ms post bounce,
the PNS radii contract rapidly to about 80~km at about 50~ms post
bounce. Using SHEN, the PNS has a smaller maximum radius of 85~km at
about 20~ms post bounce. Later, the PNS contracts on a longer
timescale on the order of several 100~ms for all EOS, where initially
SHEN leads to the smallest PNS radius up to about 200~ms post bounce.
After that, the PNS using LS180 contracts fastest and leads to the
smallest radius. LS375 has the largest PNS radius during the entire
post-bounce evolution. Thus, the PNS radius, which is often used as an
indicator of the energy budget available from core collapse, is not a
good predictor for the explosion. The same is equally true for the
compactness ($GM_\mathrm{NS}/c^2R_\mathrm{NS}$).

In stead, we conjecture that the time derivative of PNS radius would
be more important, i.e., the faster contraction of PNS is better for
the shock revival. This is consistent with the discussion of
\cite{jank12}, in which he argued that the smaller PNS radius is
better, not faster contraction. It should be noted that in his Figure
4 one finds that the model with smaller PNS radius is identical to the
faster contracting model. Therefore, our conjecture (faster
contraction is better for explosion) is valid for their simulations.

\begin{figure}[tbp]
\includegraphics[width=0.45\textwidth]{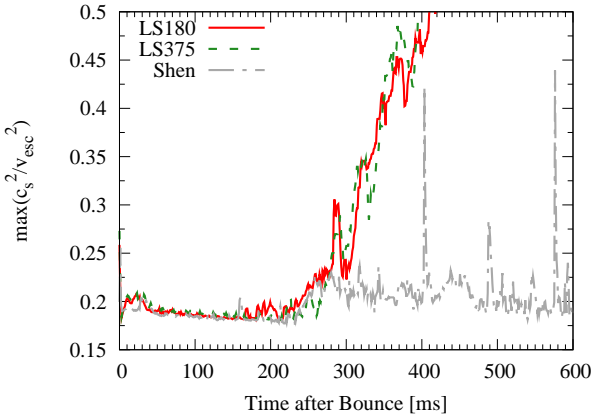}
\caption{Maximum ratio of the sound speed and the squared escape
  velocity as a function of time after bounce. The models with a
  continuously expanding shock (LS180, LS375) satisfy the criterium
  max($c_\mathrm{s}^2/v_\mathrm{esc}^2)\gtrsim 0.2$, while SHEN shows
  a very low value with the exception of few transient spikes. The
  critical value in this case should be raised to $\lesssim 0.25$.}
\label{fig:antesonic}
\end{figure}

In Figure~\ref{fig:antesonic}, we show the ante-sonic condition
proposed by \citet{pejc12}, who suggested that if
$\mathrm{max}(c_\mathrm{s}^2/v_\mathrm{esc}^2)\gtrsim
0.2,$\footnote{This critical value depends on the microphysics
  (O. Pejcha, private communication).} where $c_\mathrm{s}$ is the
sound speed and $v_\mathrm{esc}$ is the escape velocity
($=\sqrt{2\Phi}$), the stalled shock {will} turn to an actively
expanding {shock front}. For the LS EOS we find a rapid increase of
max($c_\mathrm{s}^2/v_\mathrm{esc}^2$) after about 300~ms post bounce,
which corresponds well with the onset of rapid shock expansion. For
SHEN, max($c_\mathrm{s}^2/v_\mathrm{esc}^2$) oscillates around its
maximum value of about $\lesssim 0.25$ on a longer timescale on the
order of several 100~ms. Note that the strong spike of this indicator
observed for SHEN originates from the multi-dimensional fluid motions
(e.g., shock collision with the PNS surface), which is not considered
in the 1D analysis by \citet{pejc12}. The spikes, stemming from a
transient local condition, disappear soon by energy redistribution by
neutrinos as well as by convective motions. Thus, our result suggests
that transient hot spots of the antesonic condition do not suffice to
predict an explosion.  In order to have predictive power the antesonic
condition should be satisfied in a large volume as it is the case in
our 2D models based on LS EOS .

\begin{figure}[tbp]
\includegraphics[width=0.45\textwidth]{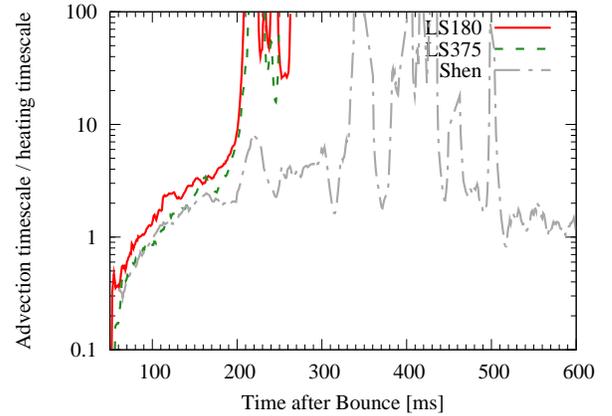}
\caption{Post-bounce evolution of the ratio of advection and heating
  timescales. The optimistic models (LS180, LS375) imply a much faster
  heating than advection through the gain region, while SHEN exhibits
  several oscillations, which correspond to the shock oscillation and
  the prolonged dwell time in the gain region.}
\label{fig:timescale}
\end{figure}

The ratio of the advection timescale ($\tau_\mathrm{adv}$) and heating
timescale ($\tau_\mathrm{heat}$) is also a well-discussed indicator of
the explosion \citep[e.g.,][]{thomp05,bura06,mare09}. This ratio is
depicted in Figure~\ref{fig:timescale}. If this indicator is greater
than unity, i.e., $\tau_\mathrm{adv}/\tau_\mathrm{heat}\gtrsim 1$, the
neutrino heating proceeds fast enough to gravitationally unbind the
fluid element, otherwise the matter is swallowed by the cooling region
below the gain radius before being heated up. Note that
$\tau_\mathrm{adv}$ is determined by the travelling time of mass
shells between shock and gain radii. This is an ``effective advection
timescale''. In 2D simulation, the velocity distribution is stochastic
because of the turbulent motion and the SASI activity so that the
simple estimate by the velocity profile and radius between shock and
gain radii is misleading. Considering a certain mass shell with
enclosed mass $M_i(r)\equiv \int \rho dV_r$, where $V_r$ is the volume
of a sphere with a radius of $r$, we note the transit time at the
shock and gain radii as $t_\mathrm{shock}(M_i)\equiv
t(M_i(r_\mathrm{shock}))$ and $t_\mathrm{gain}(M_i)\equiv
t(M_i(r_\mathrm{gain}))$, where $r_\mathrm{shock}$ and
$r_\mathrm{gain}$ mark the shock and gain radii at some time. Then, we
determine the advection timescale as $\tau_\mathrm{adv}(M_i)\equiv
t_\mathrm{gain}(M_i)-t_\mathrm{shock}(M_i)$. The angular averaged
values for the shock and gain radii are used. Moreover,
$\tau_\mathrm{heat}$ is estimated by the mass weighted integral of
$e_\mathrm{bind}/Q_\nu$ in the gain region, where $e_\mathrm{bind}$ is
the local specific binding energy (the sum of internal, kinetic and
gravitational energies) and $Q_\nu$ is the specific heating rate by
neutrinos. Figure~\ref{fig:timescale} implies that the models with
continuous and rapid shock expansion (LS180, LS375) exhibit a more
rapid increase of the ratio. For SHEN, the ratio shows also large
oscillations with several peaks (even as large as 100). However, this
model represents the least optimistic case for possible explosions and
hence the criteria of the timescale ratio is not a robust indicator
for a successful explosion.

\begin{figure*}[tbp]
\centering
\subfigure[Heating efficiency.]{
\includegraphics[width=0.475\textwidth]{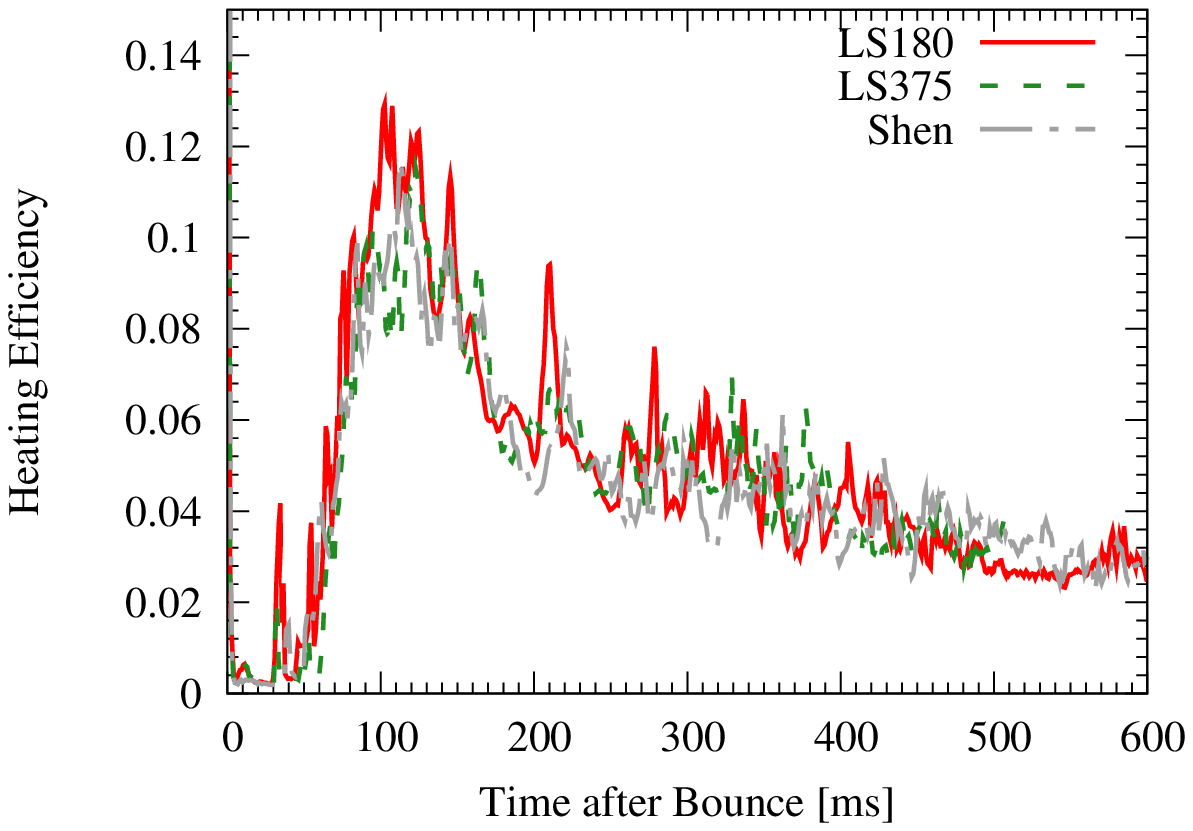}
\label{fig:heat_effciency}}
\hfill
\subfigure
[Growth parameter of convection.]{
\includegraphics[width=0.475\textwidth]{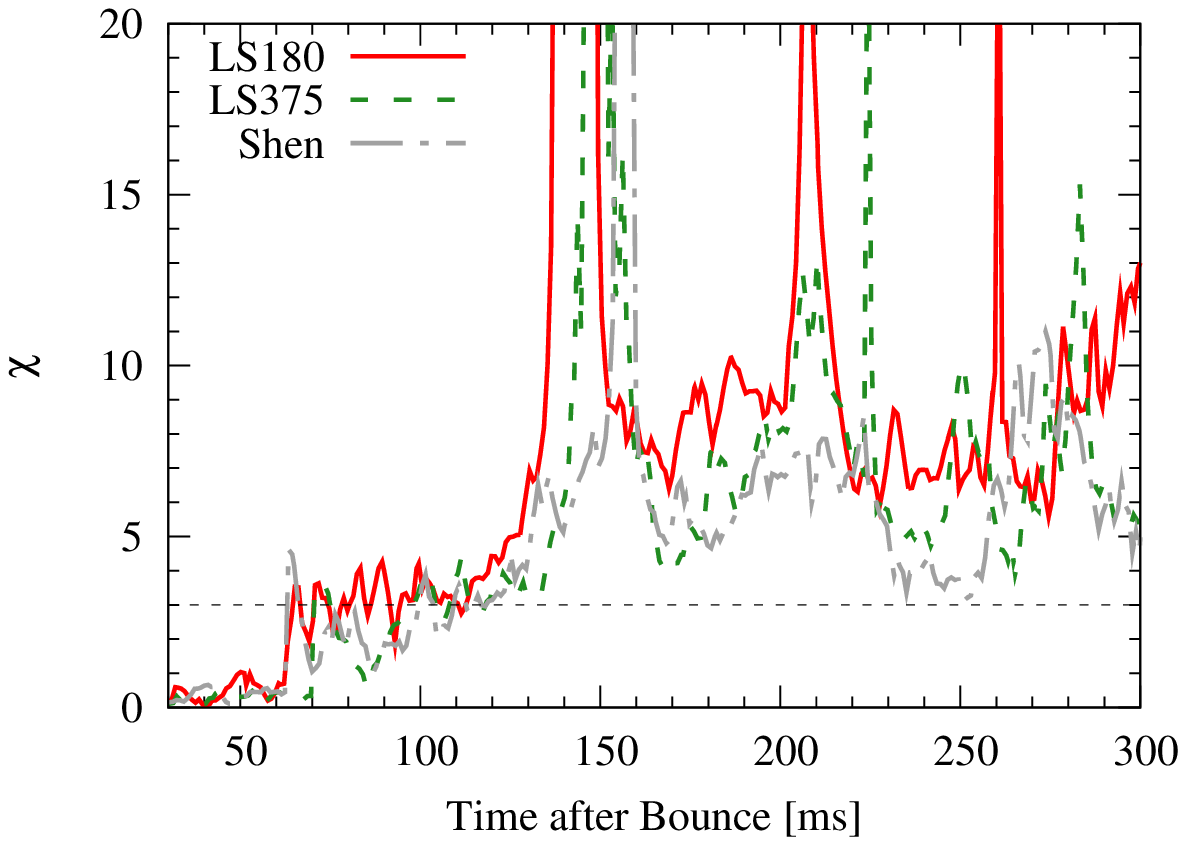}
\label{fig:chi}}
\caption{Post-bounce evolution of the heating efficiency (left) and
  the growth parameter of convection (right). In the right panel, the
  dotted horizon line represents the critical value of 3.  Both
  quantities do not exhibit the significant EOS dependence.}
\end{figure*}

Finally, the heating efficiency shown in
Figure~\ref{fig:heat_effciency} is used as a measure for the neutrino
heating process \citep{mare09,muel12b}. The definition of the heating
efficiency is given by
\begin{equation}
\eta_\mathrm{heat}=\frac{Q_\mathrm{heat}}{L_{\nu_e}+L_{\bar\nu_e}},
\label{eq:heat}
\end{equation}
where $Q_\mathrm{heat}$ is the volume-integrated neutrino heating rate
in the gain region.  We did not find any significant difference for
this indicator between models with different EOS.
In addition, we show the time evolution of the growth parameter of
convection, $\chi$, which is used as an indicator for the onset of the
convection induced by the entropy gradient
\citep{fogl06,fern09,muel12c}, defined as
\begin{equation}
\chi=\int^{r_\mathrm{shock}}_{r_\mathrm{gain}}\frac{\omega_\mathrm{BV}}{|v_{r}|}dr,
\end{equation}
where $\omega_\mathrm{BV}$ is the Brunt-V\"ais\"al\"a
frequency. \cite{fogl06} showed that $\chi$ must exceed $\sim 3$ for
convective motion to glow. Figure \ref{fig:chi} exhibits the time
evolution of $\chi$ for investigated models and suggests that EOS does
not significantly affect the onset of the convection induced by the
neutrino heating.

At this point, we remind the reader again that in this study we omit
important cooling and thermalization contributions, i.e. heavy-lepton
neutrinos and scattering on electrons/positrons.  However, these
simplifications affect the evolution of {\it all} simulations
equally. Nevertheless, we find most optimistic conditions for the
onset of an explosion in any case, in particular for the 15 $M_\odot$,
using LS180 and LS375.  Whether this remains true for LS including
these yet missing contributions, will have to be shown in a future
exploration. On the other hand, such optimistic conditions were not
obtained in $\sim $600 ms postbounce for our 2D simulations using
SHEN, even without the above mentioned important sources of energy
loss. It indicates that it may be much more difficult to obtain
neutrino-driven explosions for this progenitor model using SHEN EOS.

\subsection{Results of a model with 11.2 $M_\odot$}

\begin{figure*}[tbp]
\centering
\subfigure[Average shock radius.]{
\includegraphics[width=0.475\textwidth]{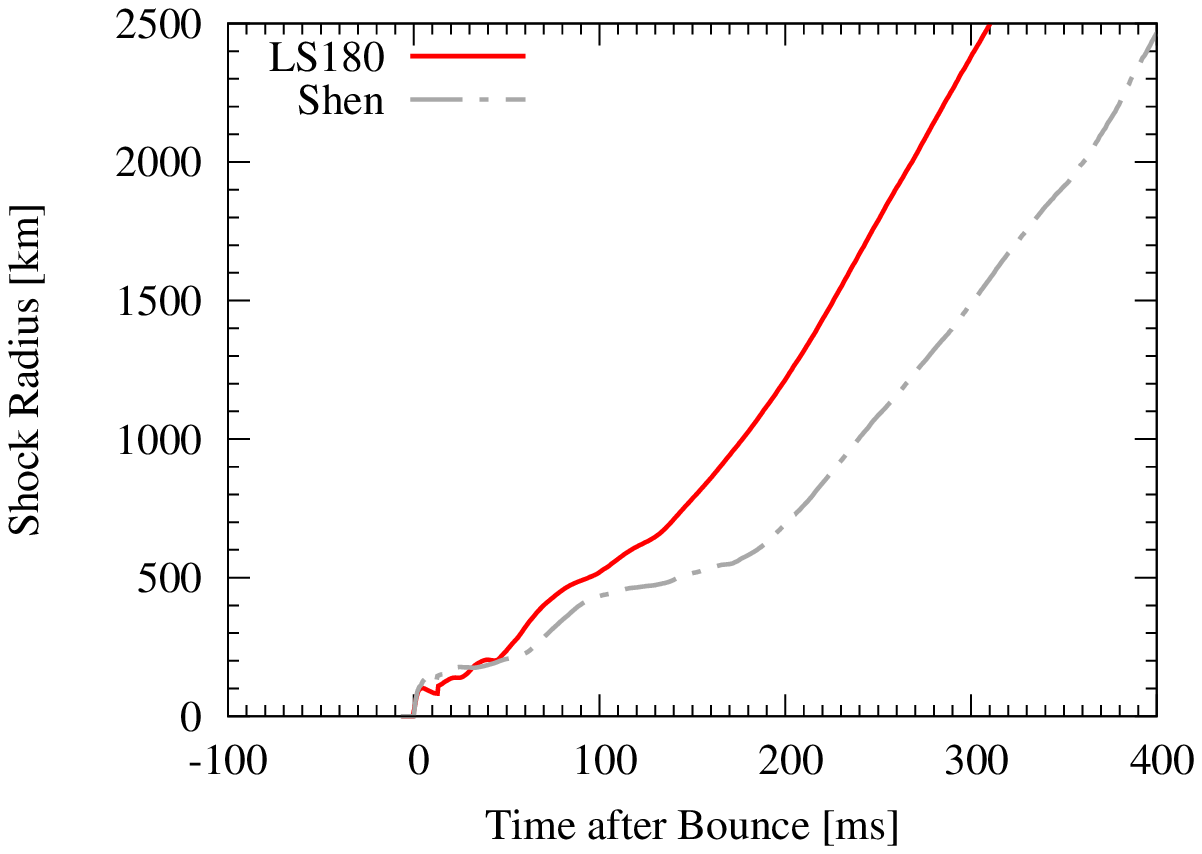}
\label{fig:shock_s112}}
\hfill
\subfigure
[The mass trajectories.]{
\includegraphics[width=0.475\textwidth]{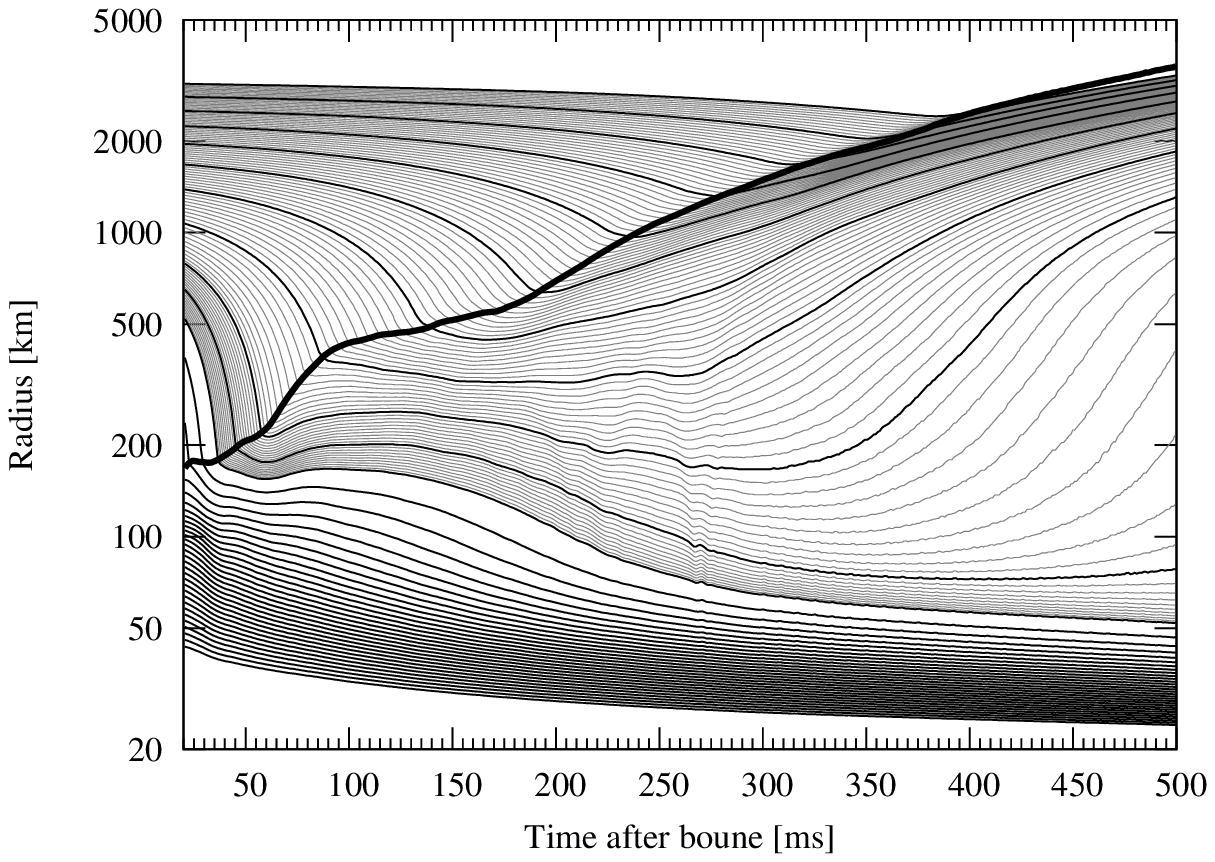}
\label{fig:mass_s112}}
\caption{Post-bounce evolution of the average shock radius (left) and
  selected mass elements (right) for a 2D simulation of a
  11.2~$M_\odot$ star. In the right panel, the thick black line is the
  angle-averaged shock radius. The thin black and grey lines show mass
  elements with an enclosed mass from 1.0 to 1.4 $M_\odot$ (thin
  black) and from 1.3 to 1.4 $M_\odot$ (grey) at intervals of 0.01
  $M_\odot$ (thin black) and 0.001 $M_\odot$ (grey), respectively.}
\end{figure*}

In addition to the 15 $M_\odot$ progenitor from \cite{woos95}, we
performed 2D simulations of an 11.2 $M_\odot$ progenitor from
\cite{woos02}.  The 11.2 $M_{\odot}$ star has been used in several
studies before, where neutrino-driven explosions were obtained in 2D
\citep{bura06,mare09} and 3D \citep{taki12} simulations.  Figure
\ref{fig:shock_s112} shows the time evolution of the average shock
radius in our 2D models, comparing LS180 (red solid line) and SHEN
(grey dot-dashed line).  For the 15 $M_\odot$ progenitor model
discussed above, only a passive shock expansion was obtained due to
the decreasing ram pressure of LS EOS. In contrast to this, both EOS
lead to the successful launch of an explosion in the case of the 11.2
$M_\odot$ progenitor, for which the expansion of the shock is
accompanied by mass outflow. This is illustrated in
Figure~\ref{fig:mass_s112}, which shows the post-bounce evolution of
selected mass shells for the simulation using SHEN. This Figure should
be compared with Figure~\ref{fig:mass_ww15}, which shows the 15
$M_{\odot}$ model using LS180. The differences are caused by the
different progenitor structures of the 15 $M_\odot$ and the 11.2
$M_\odot$ progenitor models.  The latter has a much steeper density
gradient at the interface between the iron core and the Si-layer, as
well as between the Si-layer and the C/O-layer. A more detailed
analysis regarding the criteria for the successful launch of a
neutrino-driven explosion has already been given in \cite{bura06},
where the same two progenitor stars have been compared.

\begin{figure}[tbp]
\includegraphics[width=0.45\textwidth]{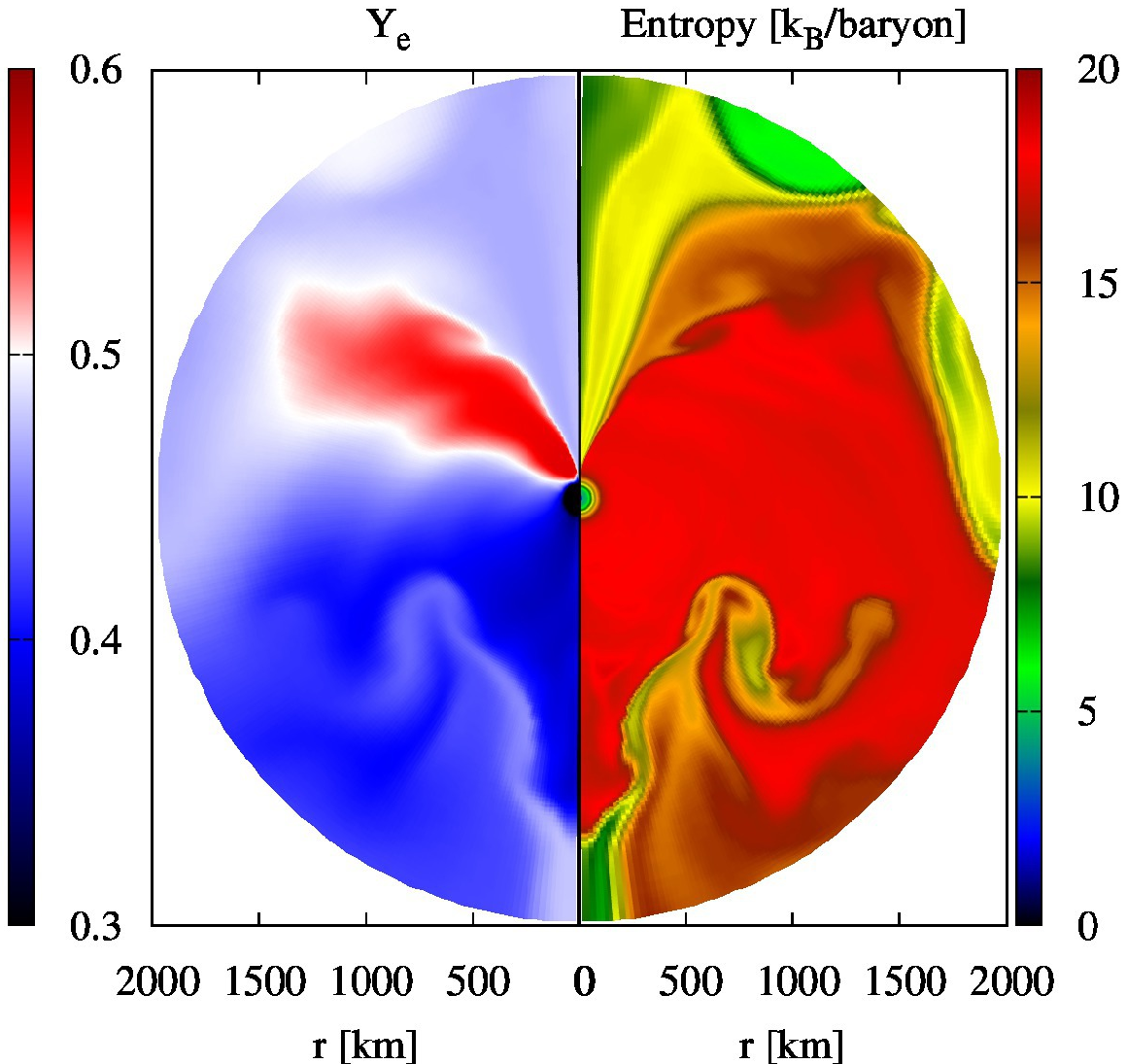}
\caption{The $Y_e$ (left panel) and entropy per baryon (right panel)
  distributions of a 2D simulation using SHEN. Shown is a snapshot at
  the post-bounce time oft 555 ms, when the model has already entered
  the neutrino-driven wind phase.}
\label{fig:s112_sye}
\end{figure}

For both, SHEN and LS180, the post-bounce accretion phase lasts for
only about 50~ms.  After this short accretion phase, the shock waves
start to expand continuously to larger radii. This is significantly
earlier than in the simulations reported by \citet{mare09,muel12b}.
In the latter, the explosion was delayed by several 100~ms. Our models
are not taking general relativistic effects into account and neglect
the emission of $\mu-$ and $\tau$-neutrinos. Both effects contribute
toward a larger radius of the shock in the post-bounce phase compared
to models with standard input physics. Moreover, the findings
described above in the comparison between SHEN and LS models of the 15
$M_\odot$ progenitor are seen in the 11.2 $M_\odot$ model as well. The
models based on SHEN lead to less favorable conditions for the
explosion. The difference between SHEN and LS180 in the 11.2 $M_\odot$
model reduces to a slight delay in the onset time of the explosion and
a slower shock expansion.  As discussed above, it is due to less
efficient neutrino heating in the SHEN case.  Nevertheless, both
explosions are weak. At the end of the simulations, the diagnostic
energy is $E_\mathrm{dia}\approx 10^{50}$~erg.

A snapshot of the $Y_e$ and entropy per baryon distributions is shown
in Figure~\ref{fig:s112_sye} at 555~ms post bounce. At this this is
the last candidate.  next esc will revert to uncompleted text. ime the
model has already entered the neutrino-driven wind phase, in which
regions of constant entropy per baryon emerge around the protoneutron
star (see fig.~\ref{fig:s112_sye}). Based on spherically symmetric
simulations of massive iron-core collapse and explosions,
\citet{fisc10,fisc11} confirmed recently the expectation of
\citet{qian96}, that the neutrino-driven wind attains proton-rich
conditions. Our models are consistent with these results: During the
early wind phase we obtain $Y_e \sim 0.5-0.55$ in the ejected matter.

\section{Summary and Discussion}\label{sec:summary}

We performed 1D and 2D radiation hydrodynamic simulations that include
spectral neutrino-radiation transfer. The simulations are launched
from a 15 $M_\odot$ progenitor star using four different EOS,
corresponding to three different incompressibilities of the LS EOS and
one SHEN EOS. LS is based on a non-relativistic approach and has a
symmetry energy of 29.3~MeV. The incompressibility parameter can be
set to 180 (LS180), 220 (LS220), or 375 MeV (LS375).  SHEN is based on
relativistic mean field theory and the Thomas-Fermi approximation. It
has a symmetry energy of 36.9 MeV and an incompressibility of 281
MeV. This set of of EOS represents a wide range of nuclear matter
properties.

In 1D, none of the simulations produced an explosion for the
considered simulation times up to 1 s post bounce. The observed
similarities between LS180, LS220 and LS375, as well as differences to
SHEN, concerning the conditions at bounce, the propagation of the
shock, and the evolution of neutrino luminosities and mean energies,
are in agreement with previous studies.  The incompressibility and
symmetry energy of the nuclear EOS depend on the density and determine
the conditions from which the explosion may emerge. However, as
recently discussed in \cite{stei12}, it is not possible to reduce the
characteristics of the EOS to a single important parameter.  Note that
there are several constraints on the nuclear properties
\citep[e.g.,][]{stei10, latt12}.  These suggest that, among the
employed EOS in this study, SHEN has too high symmetry energy, LS180
has too low incompressibility, and LS375 has too high
incompressibility.  Only LS220 is compatible with terrestrial
experiments and the 1.97 $M_\odot$ Demorest~\emph{et~al.} pulsar mass
constraint for cold neutron star matter.

In 2D, the differences between models with LS and SHEN EOS are
consistent with the differences between the 1D models. For example, we
find significantly more favorable conditions for the launch of a
neutrino-driven explosion in the models using the LS EOS: The shock
expands continuously to large radii, which are not reached in the
models using SHEN at comparable times.  We analyzed these different
post-bounce evolutions and found that models with LS show a highly
turbulent velocity field producing prompt convection during the early
post-bounce phase. Moreover, LS leads to to higher electron
antineutrino luminosity and more efficient neutrino heating. More mass
is accumulated in the gain region, which leads to a higher entropy per
baryon at the early post-bounce phase. In addition, the standing
accretion shock instability (SASI) pushes the standing accretion shock
to increasingly larger radii after a post-bounce time on the order of
100~ms. In contrast to these models based on LS, the models based on
SHEN show weak neutrino heating and the standing accretion shock is
only oscillating around a mean radius for simulation times up to
600~ms post bounce. Note, however, that our simulations ignore
inelastic scattering on electrons, which is important during the
collapse phase, and the emission of heavy-lepton neutrinos, which
contributes to the energy loss during the post-bounce phase. These
reactions would both make our models more pessimistic, so that the 2D
models based on SHEN appear unlikely to produce an explosion with the
complete set of neutrino reactions. Hence, the different properties of
the nuclear EOS can explain some apparently contradictory results
found in previous studies, for example the outcome of neutrino-driven
explosions in simulations of \cite{bura06,brue09,suwa10} using LS EOS
and the absence of neutrino-driven explosions in simulations of
\cite{burr06} using SHEN. Very recently, \cite{couc12} investigated
the EOS dependence based on simulations with parameterized neutirno
luminosities similar to \cite{murp08,nord10,hank12}.  He found that a
15 $M_{\odot}$ star is more difficult to explode using SHEN than LS
EOS, which is also consistent with the result of this paper.

The most optimistic model, with respect to a possible neutrino-driven
explosion, was obtained using LS180, for which we explore different
indicators for the possible onset of an explosion. We note again that
the neutrino-driven shock revival occurres much faster in this model
compared to \citet{mare09} because of the aforementioned neglect of
the additional cooling processes. The energy for the shock expansion
stems from aspherical mass accretion from large radii deep onto the
surface of the protoneutron star, generating a semi-stationary
advection cycle. Thus, matter which expands behind the shock wave
experiences a continuous flux of neutrinos from the accreted
matter. It remains to be shown whether the asymmetry of the accretion
pattern is similarly pronounced in three-dimensional supernova
simulations. A small difference between the $\nu_e$ and $\bar\nu_e$
spectra leads to proton-rich conditions.  Whether these findings will
remain after having included a treatment of charged-current processes
that is consistent with the EOS, as discussed recently in
\cite{mart12} and \cite{robe12}, remains to be shown in future
studies. Such improvements may become important especially at later
post-bounce times that we have not reached in the present simulations.

Based on the described explosion model we evaluate different
indicators that have been suggested in the literature to diagnose the
onset of an explosion during the post-bounce phase. We confirm that
mass-weighted average entropy in the gain region \citep{murp08} and
the mass enclosed in the gain region itself \citep{jank01} both show
the expected difference between the optimistic LS and the pessimistic
SHEN models. With respect to the ante-sonic condition \citep{pejc12},
we find that it applies as well. However, fast convection can lead to
transient peaks in the indicator that may not be interpreted as a
possible onset of the explosion. Also the ratio between advection time
scale and heating time scale has been suggested as explosion indicator
\citep{thomp05}.  This indicator is also affected by multidimensional
effects and one would have to carefully specify how the time scales
have to be averaged in a multidimensional setting. The evaluation of
the time scales based on angularly integrated quantities does not lead
to a robust indicator for the onset of an explosion. Also the
evolution of the radius of the protoneutron star does not serve as a
reliable predictor of the explosion.

In order to extend our EOS comparison study to a broader range of
initial progenitor models, and in particular to confirm the optimistic
results obtained using LS, we also include a lower mass iron-core
progenitor of 11.2 $M_\odot$ into our investigation. For this model,
we obtain neutrino-driven explosions for both LS and SHEN EOS. The
differences observed between both simulations remain qualitatively
similar to the ones discussed above for the 15 $M_\odot$ models. In
comparison to LS, SHEN leads to a slightly delayed onset of the
explosion with a smaller explosion energy.

However, it should be noted that the simulations in this paper are
only a very first step towards more realistic supernova models. In
addition to the simplifications of weak processes and the omission of
heavy-lepton neutrinos, general relativistic effects should be
considered. Additionally, the ray-by-ray approximation may lead to an
overestimation of the directional dependence of neutrino anisotropies
and of radial oscillations of SASI. A full-angle transport will give
us a more refined answer \citep[see][]{ott08,bran11,sumi12}. Moreover,
due to the coordinate symmetry axis, the SASI develops preferentially
along this axis. It could thus provide more favorable condition for
the growth of $\ell = 1$ mode of the SASI and also for the possible
onset of explosions along this direction. In the appendix we point to
the clear and crucial impact the angular width of the computational
domain has on the activity of the SASI, and in consequence on the
presence or absence of the explosion.  Hence, supernova modelers are
gradually switching gears from 2D to 3D simulations
\citep[e.g.,][]{iwak08,sche08,kota09,sche10,nord10,wang10,taki12,hank12,burrows12},
in which the SASI and convection were observed to develop much more
stochastically in 3D than in 2D. This shows that complementary 3D
supernova models are indeed necessary to pin down the impact of the
EOS on the neutrino-driven mechanism. Together with the 3D effects, it
would also be interesting to study the possible EOS impacts on the
gravitational-wave signature and neutrino emission (e.g.,
\citealt{mare09b}, \citealt{muel12}, see \citealt{kota12} for a recent
review). We will address these questions in forthcoming papers.

\acknowledgements 

We thank S. Couch, R. Fern\'andez, M. Hempel, H.-Th. Janka,
B. M\"uller, J. Murphy, T. Muto, Y. Sekiguchi, K. Sumiyoshi,
H. Suzuki, and S. Yamada for stimulating discussions.  Numerical
computations were in part carried on Cray XT4 and medium-scale
clusters at CfCA of the National Astronomical Observatory of Japan,
and on SR16000 at YITP in Kyoto University. This study was supported
in part by the Grants-in-Aid for the Scientific Research from the
Ministry of Education, Science and Culture of Japan (Nos. 19540309,
20740150, 23540323, and 23840023), MEXT HPCI STRATEGIC PROGRAM, the
Swiss National Science Foundation under grant (Nos. PP00P2-124879,
200020-132816 and PBBSP2-133378), the HP2C project 'Supernova' and HIC
for FAIR.

\appendix

\section{Importance of SASI}

\begin{figure}[tbp]
\includegraphics[width=0.45\textwidth]{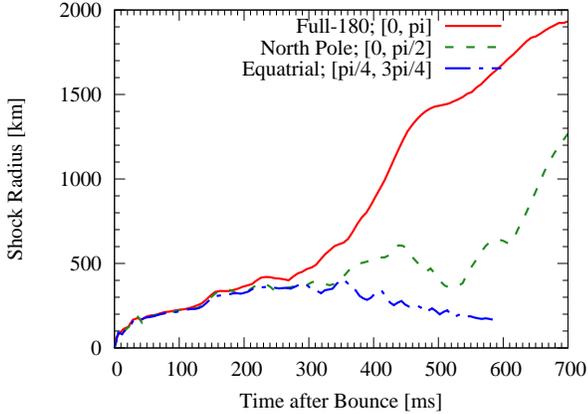}
\caption{Post-bounce evolution of the average shock radius for
  different mesh configuration.}
\label{fig:shock_mesh}
\end{figure}

In addition to the full $\pi$ simulation, we perform simulations with
$0^\circ\le\theta\le 90^\circ$ and $45^\circ\le\theta\le 135^\circ$
for LS180, which shows SASI activity. The latter grid setting was
often used in the past, cf., \citet{burr95,frye99,bura06}. The SASI is
thought to be fastest growing for low-$\ell$ modes, such as $\ell=$1
and 2 \citep{blon03}. The restriction to $0^\circ\le\theta\le
90^\circ$ does not include $\ell=1$, but it includes the $\ell=2$
mode, while the restriction to $45^\circ\le\theta\le 135^\circ$ does
include none of them. Figure~\ref{fig:shock_mesh} shows the
post-bounce evolution of the shock average radius for different
simulations using LS180 and the three above mentioned resolutions. The
red-solid, green-dotted, and blue-dashed lines indicate the full $\pi$
simulation, $\pi/2$ for the northern pole, and $\pi/2$ around the
equatorial plane, respectively. We find that the simulations with
$0^\circ\le\theta\le 90^\circ$ represent an optimistic case with
respect to possible explosions with continuous shock expansion to
larger radii. For $45^\circ\le\theta\le 135^\circ$, the shock wave
does not continue to expand to larger radii and represents the least
optimistic scenario with respect to possible explosions.  For the
latter case, where the development of possible SASI activity is
suppressed due to the chosen angular resolution, possible explosions
cannot be expected.  Thus, large degrees of freedom with respect to
angular resolution, allowing for the evolution of SASI, is also an
important requirement being able to make predictions about the
possible onset of an explosion.

\begin{thebibliography}{}

\bibitem[{{Baron} {et~al.}(1985){Baron}, {Cooperstein}, \&
    {Kahana}}]{baro85} {Baron}, E., {Cooperstein}, J., \& {Kahana},
  S. 1985, Physical Review Letters, 55, 126

\bibitem[{{Bethe}(1990)}]{beth90} {Bethe}, H.~A. 1990, Reviews of
  Modern Physics, 62, 801

\bibitem[{{Blondin} {et~al.}(2003){Blondin}, {Mezzacappa}, \&
    {DeMarino}}]{blon03} {Blondin}, J.~M., {Mezzacappa}, A., \&
  {DeMarino}, C. 2003, \apj, 584, 971

\bibitem[{{Brandt} {et~al.}(2011){Brandt}, {Burrows}, {Ott}, \&
    {Livne}}]{bran11} {Brandt}, T.~D., {Burrows}, A., {Ott}, C.~D., \&
  {Livne}, E. 2011, \apj, 728, 8

\bibitem[{{Bruenn}(1985)}]{brue85} {Bruenn}, S.~W. 1985, \apjs, 58,
  771

\bibitem[{{Bruenn} \& {Mezzacappa}(1994)}]{brue94} {Bruenn}, S.~W., \&
  {Mezzacappa}, A. 1994, \apjl, 433, L45

\bibitem[{{Bruenn} {et~al.}(2009){Bruenn}, {Mezzacappa}, {Hix},
    {Blondin}, {Marronetti}, {Messer}, {Dirk}, \& {Yoshida}}]{brue09}
  {Bruenn}, S.~W., {Mezzacappa}, A., {Hix}, W.~R., {Blondin}, J.~M.,
  {Marronetti}, P., {Messer}, O.~E.~B., {Dirk}, C.~J., \& {Yoshida},
  S. 2009, in American Institute of Physics Conference Series,
  Vol. 1111, American Institute of Physics Conference Series,
  ed. {G.~Giobbi, A.~Tornambe, G.~Raimondo, M.~Limongi,
    L.~A.~Antonelli, N.~Menci, \& E.~Brocato}, 593--601

\bibitem[{{Buras} {et~al.}(2006{\natexlab{a}}){Buras}, {Janka},
    {Rampp}, \& {Kifonidis}}]{bura06} {Buras}, R., {Janka}, H.,
  {Rampp}, M., \& {Kifonidis}, K. 2006{\natexlab{a}}, \aap, 457, 281

\bibitem[{{Buras} {et~al.}(2006{\natexlab{b}}){Buras}, {Rampp},
    {Janka}, \& {Kifonidis}}]{bura06a} {Buras}, R., {Rampp}, M.,
  {Janka}, H.-T., \& {Kifonidis}, K.  2006{\natexlab{b}}, \aap, 447,
  1049

\bibitem[{{Burrows} {et~al.}(1995){Burrows}, {Hayes}, \&
    {Fryxell}}]{burr95} {Burrows}, A., {Hayes}, J., \& {Fryxell},
  B.~A. 1995, \apj, 450, 830

\bibitem[{{Burrows} {et~al.}(2006){Burrows}, {Livne}, {Dessart},
    {Ott}, \& {Murphy}}]{burr06} {Burrows}, A., {Livne}, E.,
  {Dessart}, L., {Ott}, C.~D., \& {Murphy}, J. 2006, \apj, 640, 878

\bibitem[Burrows et al.(2012)]{burrows12} Burrows, A., Dolence, J.~C.,
  \& Murphy, J.~W.\ 2012, arXiv:1204.3088

\bibitem[Couch(2012)]{couc12} Couch, S.~M.\ 2012, arXiv:1206.4724

\bibitem[Dessart et al.(2006)]{dessart06} Dessart, L., Burrows, A.,
  Livne, E., \& Ott, C.~D.\ 2006, \apj, 645, 534

\bibitem[{{Demorest} {et~al.}(2010){Demorest}, {Pennucci}, {Ransom},
    {Roberts}, \& {Hessels}}]{demo10} {Demorest}, P.~B., {Pennucci},
  T., {Ransom}, S.~M., {Roberts}, M.~S.~E., \& {Hessels},
  J.~W.~T. 2010, \nat, 467, 1081

\bibitem[Fern{\'a}ndez \& Thompson(2009)]{fern09} Fern{\'a}ndez, R.,
  \& Thompson, C.\ 2009, \apj, 697, 1827

\bibitem[{{Fischer} {et~al.}(2010){Fischer}, {Whitehouse},
    {Mezzacappa}, {Thielemann}, \& {Liebend{\"o}rfer}}]{fisc10}
  {Fischer}, T., {Whitehouse}, S.~C., {Mezzacappa}, A., {Thielemann},
  F., \& {Liebend{\"o}rfer}, M. 2010, \aap, 517, A80+

\bibitem[{{Fischer} {et~al.}(2009){Fischer}, {Whitehouse},
    {Mezzacappa}, {Thielemann}, \& {Liebend{\"o}rfer}}]{fisc09}
  {Fischer}, T., {Whitehouse}, S.~C., {Mezzacappa}, A., {Thielemann},
  F.-K., \& {Liebend{\"o}rfer}, M. 2009, \aap, 499, 1

\bibitem[{{Fischer} {et~al.}(2011){Fischer}, {Sagert}, {Pagliara},
    {Hempel}, {Schaffner-Bielich}, {Rauscher}, {Thielemann},
    {K{\"a}ppeli}, {Mart{\'{\i}}nez-Pinedo}, \&
    {Liebend{\"o}rfer}}]{fisc11} {Fischer}, T., {et~al.} 2011, \apjs,
  194, 39

\bibitem[Foglizzo et al.(2006)]{fogl06} Foglizzo, T., Scheck, L., \&
  Janka, H.-T.\ 2006, \apj, 652, 1436

\bibitem[{{Foglizzo} {et~al.}(2007){Foglizzo}, {Galletti}, {Scheck},
    \& {Janka}}]{fogl07} {Foglizzo}, T., {Galletti}, P., {Scheck}, L.,
  \& {Janka}, H. 2007, \apj, 654, 1006

\bibitem[{{Fr{\"o}hlich} {et~al.}(2006){Fr{\"o}hlich},
    {Mart{\'{\i}}nez-Pinedo}, {Liebend{\"o}rfer}, {Thielemann},
    {Bravo}, {Hix}, {Langanke}, \& {Zinner}}]{froh06} {Fr{\"o}hlich},
  C., {Mart{\'{\i}}nez-Pinedo}, G., {Liebend{\"o}rfer}, M.,
  {Thielemann}, F.-K., {Bravo}, E., {Hix}, W.~R., {Langanke}, K., \&
  {Zinner}, N.~T. 2006, Physical Review Letters, 96, 142502

\bibitem[{{Fryer} {et~al.}(1999){Fryer}, {Benz}, {Herant}, \&
    {Colgate}}]{frye99} {Fryer}, C., {Benz}, W., {Herant}, M., \&
  {Colgate}, S.~A. 1999, \apj, 516, 892

\bibitem[Furusawa et al.(2011)]{furusawa} Furusawa, S., Yamada, S.,
  Sumiyoshi, K., \& Suzuki, H.\ 2011, \apj, 738, 178

\bibitem[Hanke et al.(2012)]{hank12} Hanke, F., Marek, A., M{\"u}ller,
  B., \& Janka, H.-T.\ 2012, \apj, 755, 138

\bibitem[Haxton(1988)]{haxton88} Haxton, W.~C.\ 1988, Physical Review
  Letters, 60, 1999

\bibitem[{{Hebeler} {et~al.}(2010){Hebeler}, {Lattimer}, {Pethick}, \&
    {Schwenk}}]{hebe10} {Hebeler}, K., {Lattimer}, J.~M., {Pethick},
  C.~J., \& {Schwenk}, A. 2010, Physical Review Letters, 105, 161102

\bibitem[{{Hempel} {et~al.}(2012){Hempel}, {Fischer},
    {Schaffner-Bielich}, \& {Liebend{\"o}rfer}}]{hemp12} {Hempel}, M.,
  {Fischer}, T., {Schaffner-Bielich}, J., \& {Liebend{\"o}rfer}, M.
  2012, \apj, 748, 70

\bibitem[{{Hempel} \& {Schaffner-Bielich}(2010)}]{hemp10} {Hempel},
  M., \& {Schaffner-Bielich}, J. 2010, Nuclear Physics A, 837, 210

\bibitem[{{Herant} {et~al.}(1992){Herant}, {Benz}, \&
    {Colgate}}]{hera92} {Herant}, M., {Benz}, W., \& {Colgate},
  S. 1992, \apj, 395, 642

\bibitem[{{Herant} {et~al.}(1994){Herant}, {Benz}, {Hix}, {Fryer}, \&
    {Colgate}}]{hera94} {Herant}, M., {Benz}, W., {Hix}, W.~R.,
  {Fryer}, C.~L., \& {Colgate}, S.~A.  1994, \apj, 435, 339

\bibitem[{{Hillebrandt} {et~al.}(1984){Hillebrandt}, {Nomoto}, \&
    {Wolff}}]{hill84} {Hillebrandt}, W., {Nomoto}, K., \& {Wolff},
  R.~G. 1984, \aap, 133, 175

\bibitem[{{Hix} {et~al.}(2003){Hix}, {Messer}, {Mezzacappa},
    {Liebend{\"o}rfer}, {Sampaio}, {Langanke}, {Dean}, \&
    {Mart{\'{\i}}nez-Pinedo}}]{hix03} {Hix}, W.~R., {Messer}, O.~E.,
  {Mezzacappa}, A., {Liebend{\"o}rfer}, M., {Sampaio}, J., {Langanke},
  K., {Dean}, D.~J., \& {Mart{\'{\i}}nez-Pinedo}, G.  2003, Physical
  Review Letters, 91, 201102

\bibitem[{{Iwakami} {et~al.}(2008){Iwakami}, {Kotake}, {Ohnishi},
    {Yamada}, \& {Sawada}}]{iwak08} {Iwakami}, W., {Kotake}, K.,
  {Ohnishi}, N., {Yamada}, S., \& {Sawada}, K. 2008, \apj, 678, 1207

\bibitem[{{Janka}(2001)}]{jank01} {Janka}, H. 2001, \aap, 368, 527

\bibitem[Janka(2012)]{jank12} Janka, H.-T.\ 2012, arXiv:1206.2503

\bibitem[{{Janka} {et~al.}(2007){Janka}, {Langanke}, {Marek},
    {Martinez-Pinedo}, \& {M{\"u}ller}}]{jank07} {Janka}, H.,
  {Langanke}, K., {Marek}, A., {Martinez-Pinedo}, G., \& {M{\"u}ller},
  B. 2007, \physrep, 442, 38

\bibitem[{{Janka} \& {Mueller}(1996)}]{jank96} {Janka}, H., \&
  {Mueller}, E. 1996, \aap, 306, 167

\bibitem[{{Kotake} {et~al.}(2006){Kotake}, {Sato}, \&
    {Takahashi}}]{kota06} {Kotake}, K., {Sato}, K., \& {Takahashi},
  K. 2006, Reports of Progress in Physics, 69, 971

\bibitem[Kotake et al.(2009)]{kota09} Kotake, K., Iwakami, W.,
  Ohnishi, N., \& Yamada, S.\ 2009, \apjl, 697, L133

\bibitem[Kotake(2011)]{kota11} Kotake, K.\ 2011, arXiv:1110.5107

\bibitem[Kotake et al.(2012)]{kota12} Kotake, K., Takiwaki, T., Suwa,
  Y., et al.\ 2012, arXiv:1204.2330

\bibitem[{{Kotake} {et~al.}(2003){Kotake}, {Yamada}, \&
    {Sato}}]{kota03} {Kotake}, K., {Yamada}, S., \& {Sato}, K. 2003,
  \prd, 68, 044023

\bibitem[{Kotake {et~al.}(2004)Kotake, Yamada, Sato, Sumiyoshi, Ono,
    \& Suzuki}]{kota04b} Kotake, K., Yamada, S., Sato, K., Sumiyoshi,
  K., Ono, H., \& Suzuki, H. 2004, \prd, 69, 124004

\bibitem[Kiuchi \& Kotake(2008)]{kiuchi08} Kiuchi, K., \& Kotake,
  K.\ 2008, \mnras, 385, 1327

\bibitem[{{Langanke} {et~al.}(2003){Langanke},
    {Mart{\'{\i}}nez-Pinedo}, {Sampaio}, {Dean}, {Hix}, {Messer},
    {Mezzacappa}, {Liebend{\"o}rfer}, {Janka}, \& {Rampp}}]{lang03}
  {Langanke}, K., {et~al.} 2003, Physical Review Letters, 90, 241102

\bibitem[Langanke et al.(2008)]{langanke08} Langanke, K.,
  Mart{\'{\i}}nez-Pinedo, G., M{\"u}ller, B., et al.\ 2008, Physical
  Review Letters, 100, 011101

\bibitem[Lattimer \& Lim(2012)]{latt12} Lattimer, J.~M., \& Lim,
  Y.\ 2012, arXiv:1203.4286

\bibitem[{{Lattimer} \& {Swesty}(1991)}]{latt91} {Lattimer}, J.~M., \&
  {Swesty}, F.~D. 1991, Nuclear Physics A, 535, 331

\bibitem[{{Liebend{\"o}rfer} {et~al.}(2004){Liebend{\"o}rfer},
    {Messer}, {Mezzacappa}, {Bruenn}, {Cardall}, \&
    {Thielemann}}]{lieb04} {Liebend{\"o}rfer}, M., {Messer}, O.~E.~B.,
  {Mezzacappa}, A., {Bruenn}, S.~W., {Cardall}, C.~Y., \&
  {Thielemann}, F. 2004, \apjs, 150, 263

\bibitem[{{Liebend{\"o}rfer} {et~al.}(2001){Liebend{\"o}rfer},
    {Mezzacappa}, {Thielemann}, {Messer}, {Hix}, \& {Bruenn}}]{lieb01}
  {Liebend{\"o}rfer}, M., {Mezzacappa}, A., {Thielemann}, F.-K.,
  {Messer}, O.~E., {Hix}, W.~R., \& {Bruenn}, S.~W. 2001, \prd, 63,
  103004

\bibitem[{{Liebend{\"o}rfer} {et~al.}(2005){Liebend{\"o}rfer},
    {Rampp}, {Janka}, \& {Mezzacappa}}]{lieb05} {Liebend{\"o}rfer},
  M., {Rampp}, M., {Janka}, H.-T., \& {Mezzacappa}, A. 2005, \apj,
  620, 840

\bibitem[{{Liebend{\"o}rfer} {et~al.}(2009){Liebend{\"o}rfer},
    {Whitehouse}, \& {Fischer}}]{lieb09} {Liebend{\"o}rfer}, M.,
  {Whitehouse}, S.~C., \& {Fischer}, T. 2009, \apj, 698, 1174

\bibitem[{{Marek} \& {Janka}(2009)}]{mare09} {Marek}, A., \& {Janka},
  H. 2009, \apj, 694, 664

\bibitem[{{Marek} {et~al.}(2009){Marek}, {Janka}, \&
    {M{\"u}ller}}]{mare09b} {Marek}, A., {Janka}, H.-T., \&
  {M{\"u}ller}, E. 2009, \aap, 496, 475

\bibitem[Mart{\'{\i}}nez-Pinedo et al.(2012)]{mart12}
  Mart{\'{\i}}nez-Pinedo, G., Fischer, T., Lohs, A., \& Huther,
  L.\ 2012, arXiv:1205.2793

\bibitem[{{M{\"u}ller} {et~al.}(2010){M{\"u}ller}, {Janka}, \&
    {Dimmelmeier}}]{muel10} {M{\"u}ller}, B., {Janka}, H., \&
  {Dimmelmeier}, H. 2010, \apjs, 189, 104

\bibitem[M{\"u}ller et al.(2012a)]{muel12b} M{\"u}ller, B., Janka,
  H.-T., \& Marek, A.\ 2012a, \apj, 756, 84

\bibitem[M{\"u}ller et al.(2012b)]{muel12c} M{\"u}ller, B., Janka,
  H.-T., \& Heger, A.\ 2012b, \apj, 761, 72

\bibitem[{{M{\"u}ller} {et~al.}(2012){M{\"u}ller}, {Janka}, \&
    {Wongwathanarat}}]{muel12} {M{\"u}ller}, E., {Janka}, H.-T., \&
  {Wongwathanarat}, A. 2012, \aap, 537, A63

\bibitem[{{Murphy} \& {Burrows}(2008)}]{murp08} {Murphy}, J.~W., \&
  {Burrows}, A. 2008, \apj, 688, 1159

\bibitem[{{Nordhaus} {et~al.}(2010){Nordhaus}, {Burrows}, {Almgren},
    \& {Bell}}]{nord10} {Nordhaus}, J., {Burrows}, A., {Almgren}, A.,
  \& {Bell}, J. 2010, \apj, 720, 694

\bibitem[{{O'Connor} \& {Ott}(2011)}]{ocon11} {O'Connor}, E., \&
  {Ott}, C.~D. 2011, \apj, 730, 70

\bibitem[Ohnishi et al.(2007)]{ohnishi08} Ohnishi, N., Kotake, K., \&
  Yamada, S.\ 2007, \apj, 667, 375

\bibitem[{{Ott} {et~al.}(2008){Ott}, {Burrows}, {Dessart}, \&
    {Livne}}]{ott08} {Ott}, C.~D., {Burrows}, A., {Dessart}, L., \&
  {Livne}, E. 2008, \apj, 685, 1069

\bibitem[Pejcha \& Thompson(2012)]{pejc12} Pejcha, O., \& Thompson,
  T.~A.\ 2012, \apj, 746, 106

\bibitem[{{Pruet} {et~al.}(2006){Pruet}, {Hoffman}, {Woosley},
    {Janka}, \& {Buras}}]{prue06} {Pruet}, J., {Hoffman}, R.~D.,
  {Woosley}, S.~E., {Janka}, H.-T., \& {Buras}, R.  2006, \apj, 644,
  1028

\bibitem[Qian \& Woosley(1996)]{qian96} Qian, Y.-Z., \& Woosley,
  S.~E.\ 1996, \apj, 471, 331

\bibitem[Roberts \& Reddy(2012)]{robe12} Roberts, L.~F., \& Reddy,
  S.\ 2012, arXiv:1205.4066

\bibitem[{{Sagert} {et~al.}(2009){Sagert}, {Fischer}, {Hempel},
    {Pagliara}, {Schaffner-Bielich}, {Mezzacappa}, {Thielemann}, \&
    {Liebend{\"o}rfer}}]{sage09} {Sagert}, I., {Fischer}, T.,
  {Hempel}, M., {Pagliara}, G., {Schaffner-Bielich}, J., {Mezzacappa},
  A., {Thielemann}, F., \& {Liebend{\"o}rfer}, M. 2009, Physical
  Review Letters, 102, 081101

\bibitem[{{Scheidegger} {et~al.}(2008){Scheidegger}, {Fischer},
    {Whitehouse}, \& {Liebend{\"o}rfer}}]{sche08} {Scheidegger}, S.,
  {Fischer}, T., {Whitehouse}, S.~C., \& {Liebend{\"o}rfer}, M. 2008,
  \aap, 490, 231

\bibitem[{{Scheidegger} {et~al.}(2010){Scheidegger}, {K{\"a}ppeli},
    {Whitehouse}, {Fischer}, \& {Liebend{\"o}rfer}}]{sche10}
  {Scheidegger}, S., {K{\"a}ppeli}, R., {Whitehouse}, S.~C.,
  {Fischer}, T., \& {Liebend{\"o}rfer}, M. 2010, \aap, 514, A51+

\bibitem[{{Shen} {et~al.}(2011){Shen}, {Horowitz}, \&
    {O'Connor}}]{gshen} {Shen}, G., {Horowitz}, C.~J., \& {O'Connor},
  E. 2011, \prc, 83, 065808

\bibitem[{Shen {et~al.}(1998)Shen, Toki, Oyamatsu, \&
    Sumiyoshi}]{shen98} Shen, H., Toki, H., Oyamatsu, K., \&
  Sumiyoshi, K. 1998, Nucl. Phys., A637, 435

\bibitem[{{Steiner} {et~al.}(2010){Steiner}, {Lattimer}, \&
    {Brown}}]{stei10} {Steiner}, A.~W., {Lattimer}, J.~M., \& {Brown},
  E.~F. 2010, \apj, 722, 33

\bibitem[Steiner et al.(2012)]{stei12} Steiner, A.~W., Hempel, M., \&
  Fischer, T.\ 2012, arXiv:1207.2184

\bibitem[{{Stone} \& {Norman}(1992)}]{ston92} {Stone}, J.~M., \&
  {Norman}, M.~L. 1992, \apjs, 80, 753

\bibitem[Sumiyoshi \& Yamada(2012)]{sumi12} Sumiyoshi, K., \& Yamada,
  S.\ 2012, \apjs, 199, 17

\bibitem[{{Sumiyoshi} {et~al.}(2007){Sumiyoshi}, {Yamada}, \&
    {Suzuki}}]{sumi07} {Sumiyoshi}, K., {Yamada}, S., \& {Suzuki},
  H. 2007, \apj, 667, 382

\bibitem[{{Sumiyoshi} {et~al.}(2006){Sumiyoshi}, {Yamada}, {Suzuki},
    \& {Chiba}}]{sumi06} {Sumiyoshi}, K., {Yamada}, S., {Suzuki}, H.,
  \& {Chiba}, S. 2006, Physical Review Letters, 97, 091101

\bibitem[{{Sumiyoshi} {et~al.}(2005){Sumiyoshi}, {Yamada}, {Suzuki},
    {Shen}, {Chiba}, \& {Toki}}]{sumi05} {Sumiyoshi}, K., {Yamada},
  S., {Suzuki}, H., {Shen}, H., {Chiba}, S., \& {Toki}, H. 2005, \apj,
  629, 922

\bibitem[{{Suwa} {et~al.}(2011){Suwa}, {Kotake}, {Takiwaki},
    {Liebend{\"o}rfer}, \& {Sato}}]{suwa11b} {Suwa}, Y., {Kotake}, K.,
  {Takiwaki}, T., {Liebend{\"o}rfer}, M., \& {Sato}, K.  2011, \apj,
  738, 165

\bibitem[{{Suwa} {et~al.}(2010){Suwa}, {Kotake}, {Takiwaki},
    {Whitehouse}, {Liebend\"orfer}, \& {Sato}}]{suwa10} {Suwa}, Y.,
  {Kotake}, K., {Takiwaki}, T., {Whitehouse}, S.~C., {Liebend\"orfer},
  M., \& {Sato}, K. 2010, \pasj, 62, L49

\bibitem[{{Suwa} {et~al.}(2007{\natexlab{a}}){Suwa}, {Takiwaki},
    {Kotake}, \& {Sato}}]{suwa07a} {Suwa}, Y., {Takiwaki}, T.,
  {Kotake}, K., \& {Sato}, K. 2007{\natexlab{a}}, \apjl, 665, L43

\bibitem[{{Suwa} {et~al.}(2007{\natexlab{b}}){Suwa}, {Takiwaki},
    {Kotake}, \& {Sato}}]{suwa07b} ---. 2007{\natexlab{b}}, \pasj, 59,
  771

\bibitem[{{Suwa} {et~al.}(2009){Suwa}, {Takiwaki}, {Kotake}, \&
    {Sato}}]{suwa09a} ---. 2009, \apj, 690, 913

\bibitem[{{Takahara} \& {Sato}(1982)}]{taka82} {Takahara}, M., \&
  {Sato}, K. 1982, Progress of Theoretical Physics, 68, 795

\bibitem[Takiwaki et al.(2004)]{taki04} Takiwaki, T., Kotake, K.,
  Nagataki, S., \& Sato, K.\ 2004, \apj, 616, 1086

\bibitem[{{Takiwaki} {et~al.}(2012){Takiwaki}, {Kotake}, \&
    {Suwa}}]{taki12} {Takiwaki}, T., {Kotake}, K., \& {Suwa}, Y. 2012,
  \apj, 749, 98

\bibitem[Thompson et al.(2005)]{thomp05} Thompson, T.~A., Quataert,
  E., \& Burrows, A.\ 2005, \apj, 620, 861

\bibitem[{{Thompson} {et~al.}(2003){Thompson}, {Burrows}, \&
    {Pinto}}]{thom03} {Thompson}, T.~A., {Burrows}, A., \& {Pinto},
  P.~A. 2003, \apj, 592, 434

\bibitem[{{Wanajo}(2006)}]{wana06} {Wanajo}, S. 2006, \apj, 647, 1323

\bibitem[{{Wanajo} {et~al.}(2011){Wanajo}, {Janka}, \&
    {M{\"u}ller}}]{wana11} {Wanajo}, S., {Janka}, H.-T., \&
  {M{\"u}ller}, B. 2011, \apjl, 726, L15

\bibitem[{{Wongwathanarat} {et~al.}(2010){Wongwathanarat}, {Janka}, \&
    {M{\"u}ller}}]{wang10} {Wongwathanarat}, A., {Janka}, H.-T., \&
  {M{\"u}ller}, E. 2010, \apjl, 725, L106

\bibitem[{{Woosley} {et~al.}(2002){Woosley}, {Heger}, \&
    {Weaver}}]{woos02} {Woosley}, S.~E., {Heger}, A., \& {Weaver},
  T.~A. 2002, Reviews of Modern Physics, 74, 1015

\bibitem[{{Woosley} \& {Weaver}(1995)}]{woos95} {Woosley}, S.~E., \&
  {Weaver}, T.~A. 1995, \apjs, 101, 181


\end{thebibliography}
\end{document}